\documentclass[preprint,prd,aps,showpacs,showkeys,nofootinbib]{revtex4}
\usepackage{graphicx}
\usepackage{dcolumn}
\usepackage{bm}
\usepackage{color}
\usepackage[dvipsnames]{xcolor}
\usepackage{amssymb}

\definecolor{light-gray}{gray}{0.88}
\definecolor{dark-gray}{gray}{0.40}
\begin{document}

\title{Lepton flavor violating decays $\tau\rightarrow{Pl}$ in the $U(1)_X$SSM model}
\author{Tong-Tong Wang$^{1,2}$\footnote{wtt961018@163.com}, Shu-Min Zhao$^{1,2}$\footnote{zhaosm@hbu.edu.cn}, Xing-Xing Dong$^{1,2}$\footnote{dongxx@hbu.edu.cn}, Lu-Hao Su$^{1,2}$, Ze-Ning Zhang$^{1,2}$, Wei Li$^{1,2}$, Tai-Fu Feng$^{1,2,3}$}

\affiliation{$^1$ Department of Physics, Hebei University, Baoding 071002, China}
\affiliation{$^2$ Key Laboratory of High-precision Computation and Application of Quantum Field Theory of Hebei Province, Baoding 071002, China}
\affiliation{$^3$ Department of Physics, Chongqing University, Chongqing 401331, China}
\date{\today}

\begin{abstract}

 $U(1)_X$SSM is the extension of the minimal supersymmetric standard model(MSSM) and its local
 gauge group is $SU(3)_C\times SU(2)_L \times U(1)_Y \times U(1)_X$. To obtain this model, three singlet new Higgs superfields and right-handed neutrinos are added to MSSM.
In the framework of $U(1)_X$SSM, we study lepton flavor violating decays $\tau\rightarrow{Pl}(P=\pi,~\eta,~\eta^\prime;~l=\mu,~e)$. According to the latest experimental data of $\tau\rightarrow{Pe}$ and $\tau\rightarrow{P\mu}$, the influence of different sensitive parameters on the results is analyzed to make a reasonable prediction for future experiments. From the numerical analysis, the non-diagonal elements which correspond to the generations of the initial lepton and final lepton
are main sensitive parameters and lepton flavor violation(LFV) sources. This work can provide a basis for finding the existence of new physics(NP).

\end{abstract}


\maketitle

\section{introduction}

Neutrino oscillation experiment provides evidence of LFV\cite{neutrino}. In the standard model(SM),
the branching ratio of  lepton flavor violating process is depressed due to the very small neutrino mass,
which is very difficult to be observed in current or future experiments\cite{SMsmall}. However, for some new models based on the extension of the SM, the branching ratio of lepton flavor violating process is enhanced due to the introduction of new lepton flavor violating sources, which can be improved to the measurable range of the current experiment. Therefore, the study of LFV is very important to find NP.

 We extend the MSSM with $U(1)$ local gauge group, whose symmetry group is $SU(3)_C\times SU(2)_L \times U(1)_Y\times U(1)_X$\cite{Sarah1,Sarah2,Sarah3}. It adds three Higgs singlet superfields and right-handed neutrino superfields beyond MSSM\cite{MSSM}. In this way, light neutrinos obtain tiny masses through the seesaw mechanism, which can explain the results of neutrino oscillation experiment.
 There are five neutral CP-even Higgs component fields in the model, which come from two Higgs doublets and three Higgs singlets respectively. Therefore, the mass mixing matrix is $5\times5$, and the 125.1 GeV Higgs particle\cite{sm1} corresponds to the lightest mass eigenstate. The little hierarchy problem in MSSM is relieved in $U(1)_X$SSM by the right-handed neutrinos, sneutrinos and additional Higgs singlets. MSSM exists $\mu$ problem, while in the $U(1)_X$SSM\cite{sm2} this problem can be alleviated by the $S$ field after vacuum spontaneous breaking.

 To improve the corrections to LFV processes of $\tau \rightarrow Pl$, people extend SM in various ways.
  In the non-SUSY models including two Higgs doublet models\cite{sun1,sun2}, littlest Higgs model with T parity\cite{sun4}, TC2 model\cite{sun3}, etc.,
 there are a certain number of works for LFV processes.
  On the other hand, within SUSY models there are also some works for the LFV processes of $\tau \rightarrow Pl$.
 Some of the related SUSY models are MSSM\cite{sun5,sun6}, supersymmetric seesaw mechanism model\cite{sun7}, R-parity violating SUSY\cite{sun8}. The LFV decays $\tau \rightarrow Pl$  are also studied in the scenario
of the minimal R-symmetric supersymmetric standard model. The effective field theory is used to study $\tau \rightarrow Pl$,
and the Wilson coefficients of the LFV operators are restricted\cite{sun9}.

This work is to study the LFV of the $\tau\rightarrow{Pl}$ processes under the $U(1)_X$SSM model. The constraints from neutrino experiment, $l_j\rightarrow l_i\gamma$ and $\mu\rightarrow eee$ are taken into account.  We deduce the relevant Feynman diagrams, and numerically analyze the Feynman amplitude, decay width, and use the combination of random scattering and equal step search method to efficiently find reasonable parameter spaces. The effects of different reasonable parameter spaces on the branching ratio Br($\tau\rightarrow{Pl}$) are compared. The latest observed upper limits on the LFV branching ratios of $\tau\rightarrow{P\mu}$ and $\tau\rightarrow{Pe}$($P=\pi,\eta,\eta^\prime$) at 90\% confidence level (C.L.)\cite{sun10} are
\begin{eqnarray}
Br(\tau\rightarrow{e\pi})<8.0\times10^{-8},~~Br(\tau\rightarrow{e\eta})<9.2\times10^{-8},~~Br(\tau\rightarrow{e\eta^\prime})<1.6\times10^{-7}, \nonumber\\
Br(\tau\rightarrow{\mu\pi})<1.1\times10^{-7},~~Br(\tau\rightarrow{\mu\eta})<6.5\times10^{-8},~~Br(\tau\rightarrow{\mu\eta^\prime})<1.3\times10^{-7}.
\end{eqnarray}

The paper is organized as follows. In Sec.II, we mainly introduce the $U(1)_X$SSM including its superpotential and the general soft breaking terms. In Sec.III, we give analytic expressions for the branching ratios of $\tau\rightarrow{Pl}$ decays in the $U(1)_X$SSM. In Sec.IV, we give the numerical analysis, and the summary is given in Sec.V. Finally, some mass matrices and formulas are collected in the Appendix \ref{A1} and \ref{A2}.

\section{the $U(1)_X$SSM}

 As the U(1) extension of MSSM, the local gauge group of $U(1)_X$SSM is  $SU(3)_C\otimes
SU(2)_L \otimes U(1)_Y\otimes U(1)_X$\cite{UU1,UU3,UU4}. $U(1)_X$SSM has new superfields such as three Higgs
singlets $\hat{\eta},~\hat{\bar{\eta}},~\hat{S}$ and right-handed neutrinos $\hat{\nu}_i$,
 which are beyond MSSM. Through the seesaw mechanism, light
neutrinos obtain tiny mass at the tree level.
The neutral CP-even parts of
$H_u,~ H_d,~\eta,~\bar{\eta}$ and $S$ mix together and form $5\times5 $ mass squared matrix, whose
lightest eigenvalue corresponds to the lightest CP-even Higgs mass. To get 125.1 GeV Higgs mass\cite{LCTHiggs1,LCTHiggs2},
the loop corrections should be taken into account. The sneutrinos are disparted into CP-even sneutrinos and CP-odd sneutrinos,
and their mass squared matrixes are both extended to $6\times6$.
\begin{table}
\caption{ The superfields in $U(1)_X$SSM}
\begin{tabular}{|c|c|c|c|c|}
\hline
Superfields & $SU(3)_C$ & $SU(2)_L$ & $U(1)_Y$ & $U(1)_X$ \\
\hline
$\hat{Q}_i$ & 3 & 2 & 1/6 & 0 \\
\hline
$\hat{u}^c_i$ & $\bar{3}$ & 1 & -2/3 & -$1/2$ \\
\hline
$\hat{d}^c_i$ & $\bar{3}$ & 1 & 1/3 & $1/2$  \\
\hline
$\hat{L}_i$ & 1 & 2 & -1/2 & 0  \\
\hline
$\hat{e}^c_i$ & 1 & 1 & 1 & $1/2$  \\
\hline
$\hat{\nu}_i$ & 1 & 1 & 0 & -$1/2$ \\
\hline
$\hat{H}_u$ & 1 & 2 & 1/2 & 1/2\\
\hline
$\hat{H}_d$ & 1 & 2 & -1/2 & -1/2 \\
\hline
$\hat{\eta}$ & 1 & 1 & 0 & -1 \\
\hline
$\hat{\bar{\eta}}$ & 1 & 1 & 0 & 1\\
\hline
$\hat{S}$ & 1 & 1 & 0 & 0 \\
\hline
\end{tabular}
\label{I}
\end{table}

 In $U(1)_X$SSM, the concrete form of the superpotential is:
\begin{eqnarray}
&&W=l_W\hat{S}+\mu\hat{H}_u\hat{H}_d+M_S\hat{S}\hat{S}-Y_d\hat{d}\hat{q}\hat{H}_d-Y_e\hat{e}\hat{l}\hat{H}_d+\lambda_H\hat{S}\hat{H}_u\hat{H}_d
\nonumber\\&&~~~~~~+\lambda_C\hat{S}\hat{\eta}\hat{\bar{\eta}}+\frac{\kappa}{3}\hat{S}\hat{S}\hat{S}+Y_u\hat{u}\hat{q}\hat{H}_u+Y_X\hat{\nu}\hat{\bar{\eta}}\hat{\nu}
+Y_\nu\hat{\nu}\hat{l}\hat{H}_u.
\end{eqnarray}

We collect the explicit forms of two Higgs doublets and three Higgs singlets here
\begin{eqnarray}
&&H_{u}=\left(\begin{array}{c}H_{u}^+\\{1\over\sqrt{2}}\Big(v_{u}+H_{u}^0+iP_{u}^0\Big)\end{array}\right),
~~~~~~
H_{d}=\left(\begin{array}{c}{1\over\sqrt{2}}\Big(v_{d}+H_{d}^0+iP_{d}^0\Big)\\H_{d}^-\end{array}\right),
\nonumber\\
&&\eta={1\over\sqrt{2}}\Big(v_{\eta}+\phi_{\eta}^0+iP_{\eta}^0\Big),~~~~~~~~~~~~~~~
\bar{\eta}={1\over\sqrt{2}}\Big(v_{\bar{\eta}}+\phi_{\bar{\eta}}^0+iP_{\bar{\eta}}^0\Big),\nonumber\\&&
\hspace{4.0cm}S={1\over\sqrt{2}}\Big(v_{S}+\phi_{S}^0+iP_{S}^0\Big).
\end{eqnarray}

The vacuum expectation values(VEVs) of the Higgs superfields $H_u$, $H_d$, $\eta$, $\bar{\eta}$ and $S$
are denoted by $v_u,~v_d,~v_\eta$,~ $v_{\bar\eta}$ and $v_S$ respectively. Two angles are defined as
$\tan\beta=v_u/v_d$ and $\tan\beta_\eta=v_{\bar{\eta}}/v_{\eta}$.

The soft SUSY breaking terms of this model are shown as
\begin{eqnarray}
&&\mathcal{L}_{soft}=\mathcal{L}_{soft}^{MSSM}-B_SS^2-L_SS-\frac{T_\kappa}{3}S^3-T_{\lambda_C}S\eta\bar{\eta}
+\epsilon_{ij}T_{\lambda_H}SH_d^iH_u^j\nonumber\\&&\hspace{1.5cm}
-T_X^{IJ}\bar{\eta}\tilde{\nu}_R^{*I}\tilde{\nu}_R^{*J}
+\epsilon_{ij}T^{IJ}_{\nu}H_u^i\tilde{\nu}_R^{I*}\tilde{l}_j^J
-m_{\eta}^2|\eta|^2-m_{\bar{\eta}}^2|\bar{\eta}|^2-m_S^2S^2\nonumber\\&&\hspace{1.5cm}
-(m_{\tilde{\nu}_R}^2)^{IJ}\tilde{\nu}_R^{I*}\tilde{\nu}_R^{J}
-\frac{1}{2}\Big(M_X\lambda^2_{\tilde{X}}+2M_{BB^\prime}\lambda_{\tilde{B}}\lambda_{\tilde{X}}\Big)+h.c.
\end{eqnarray}

The particle content and charge assignments for $U(1)_X$SSM are shown in the Table \ref {I}.
In our previous work, we have proven that $U(1)_X$SSM is  anomaly free\cite{UU3}.
Different from MSSM, $U(1)_X$SSM has a new effect called as the gauge kinetic mixing,
which is produced from two Abelian groups $U(1)_Y$ and $U(1)_X$.

In general, the covariant derivatives of $U(1)_X$SSM can be written as \cite{UMSSM5,B-L1,B-L2,gaugemass}
\begin{eqnarray}
&&D_\mu=\partial_\mu-i\left(\begin{array}{cc}Y,&X\end{array}\right)
\left(\begin{array}{cc}g_{Y},&g{'}_{{YX}}\\g{'}_{{XY}},&g{'}_{{X}}\end{array}\right)
\left(\begin{array}{c}A_{\mu}^{\prime Y} \\ A_{\mu}^{\prime X}\end{array}\right)\;,
\label{gauge1}
\end{eqnarray}
 with $A_{\mu}^{\prime Y}$ and $A^{\prime X}_\mu$ representing the gauge fields of $U(1)_Y$ and $U(1)_X$ respectively.

 Based on  the two unbroken Abelian gauge groups, we can perform the following operation
  \cite{UMSSM5,B-L2,gaugemass}
  \begin{eqnarray}
&&\left(\begin{array}{cc}g_{Y},&g{'}_{{YX}}\\g{'}_{{XY}},&g{'}_{{X}}\end{array}\right)
R^T=\left(\begin{array}{cc}g_{1},&g_{{YX}}\\0,&g_{{X}}\end{array}\right)\;.
\label{gauge3}
\end{eqnarray}
Here, $R$ is the rotation matrix.

At the tree level, three neutral gauge bosons $A^{X}_\mu,~A^Y_\mu$ and $V^3_\mu$ mix together, whose mass matrix
is shown in the basis $(A^Y_\mu, V^3_\mu, A^{X}_\mu)$\cite{UU4}
\begin{eqnarray}
&&\left(\begin{array}{*{20}{c}}
\frac{1}{8}g_{1}^2 v^2 &~~~ -\frac{1}{8}g_{1}g_{2} v^2 & ~~~\frac{1}{8}g_{1}(g_{{YX}}+g_{X}) v^2 \\
-\frac{1}{8}g_{1}g_{2} v^2 &~~~ \frac{1}{8}g_{2}^2 v^2 & ~~~~-\frac{1}{8}g_{2}(g_{{YX}}+g_X) v^2\\
\frac{1}{8}g_{1}(g_{{YX}}+g_{X}) v^2 &~~~ -\frac{1}{8}g_{2}(g_{{YX}}+g_{X}) v^2 &~~~~ \frac{1}{8}(g_{{YX}}+g_{X})^2 v^2+\frac{1}{8}g_{{X}}^2 \xi^2
\end{array}\right),\label{gauge matrix}
\end{eqnarray}
with $v^2=v_u^2+v_d^2$ and $\xi^2=v_\eta^2+v_{\bar{\eta}}^2$.

To get mass eigenvalues  of the matrix in Eq.(\ref{gauge matrix}), we use
 two mixing angles $\theta_{W}$ and $\theta_{W}'$.
 $\theta_{W}$ is the Weinberg angle and the new mixing angle $\theta_{W}'$ is
 defined from the following formula
\begin{eqnarray}
\sin^2\theta_{W}'=\frac{1}{2}-\frac{[(g_{{YX}}+g_{X})^2-g_{1}^2-g_{2}^2]v^2+
4g_{X}^2\xi^2}{2\sqrt{[(g_{{YX}}+g_{X})^2+g_{1}^2+g_{2}^2]^2v^4+8g_{X}^2[(g_{{YX}}+g_{X})^2-g_{1}^2-g_{2}^2]v^2\xi^2+16g_{X}^4\xi^4}}.
\end{eqnarray}

It appears in the couplings involving $Z$ and $Z^{\prime}$.
The exact eigenvalues of Eq. (\ref{gauge matrix}) are deduced \cite{UU4}
\begin{eqnarray}
&&m_\gamma^2=0,\nonumber\\
&&m_{Z,{Z^{'}}}^2=\frac{1}{8}\Big([g_{1}^2+g_2^2+(g_{{YX}}+g_{X})^2]v^2+4g_{X}^2\xi^2 \nonumber\\
&&\hspace{1.1cm}\mp\sqrt{[g_{1}^2+g_{2}^2+(g_{{YX}}+g_{X})^2]^2v^4+8[(g_{{YX}}+g_{X})^2-g_{1}^2-
g_{2}^2]g_{X}^2v^2\xi^2+16g_{X}^4\xi^4}\Big).
\end{eqnarray}

The mass squared matrix for CP-even sneutrino $({\phi}_{l}, {\phi}_{r})$ reads
\begin{eqnarray}
M^2_{\tilde{\nu}^R} = \left(
\begin{array}{cc}
m_{{\phi}_{l}{\phi}_{l}} &m^T_{{\phi}_{r}{\phi}_{l}}\\
m_{{\phi}_{l}{\phi}_{r}} &m_{{\phi}_{r}{\phi}_{r}}\end{array}
\right),\label{Rsneu}
 \end{eqnarray}
\begin{eqnarray}
&&m_{{\phi}_{l}{\phi}_{l}}= \frac{1}{8} \Big((g_{1}^{2} + g_{Y X}^{2} + g_{2}^{2}+ g_{Y X} g_{X})( v_{d}^{2}- v_{u}^{2})
+  g_{Y X} g_{X}(2 v_{\eta}^{2}-2 v_{\bar{\eta}}^{2})\Big)
\nonumber\\&&\hspace{1.8cm}+\frac{1}{2} v_{u}^{2}{Y_{\nu}^{T}  Y_\nu}  + M_{\tilde{L}}^2,
 \\&&m_{{\phi}_{l}{\phi}_{r}} = \frac{1}{\sqrt{2} } v_uT_\nu  +  v_u v_{\bar{\eta}} {Y_X  Y_\nu}
  - \frac{1}{2}v_d ({\lambda}_{H}v_S  + \sqrt{2} \mu )Y_\nu,\\&&
m_{{\phi}_{r}{\phi}_{r}}= \frac{1}{8} \Big((g_{Y X} g_{X}+g_{X}^{2})(v_{d}^{2}- v_{u}^{2})
+2g_{X}^{2}(v_{\eta}^{2}- v_{\bar{\eta}}^{2})\Big) + v_{\eta} v_S Y_X {\lambda}_{C}\nonumber \\&&\hspace{1.8cm}
 +M_{\tilde{\nu}}^2 + \frac{1}{2} v_{u}^{2}|Y_\nu|^2+  v_{\bar{\eta}} (2 v_{\bar{\eta}}Y_X  Y_X  + \sqrt{2} T_X).
\end{eqnarray}

The matrix $Z^R$ is used to diagonalize $M^2_{\tilde{\nu}^R}$.
We also deduce the mass squared matrix for CP-odd sneutrino $({\sigma}_{l}, {\sigma}_{r})$ and
use the matrix $Z^I$ to diagonalize it,
\begin{eqnarray}
M^2_{\tilde{\nu}^I} = \left(
\begin{array}{cc}
m_{{\sigma}_{l}{\sigma}_{l}} &m^T_{{\sigma}_{r}{\sigma}_{l}}\\
m_{{\sigma}_{l}{\sigma}_{r}} &m_{{\sigma}_{r}{\sigma}_{r}}\end{array}
\right),
 \end{eqnarray}
\begin{eqnarray}
&&m_{{\sigma}_{l}{\sigma}_{l}}= \frac{1}{8} \Big((g_{1}^{2} + g_{Y X}^{2} + g_{2}^{2}+  g_{Y X} g_{X})( v_{d}^{2}- v_{u}^{2})
+  2g_{Y X} g_{X}(v_{\eta}^{2}-v_{\bar{\eta}}^{2})\Big)
\nonumber\\&&\hspace{1.8cm}+\frac{1}{2} v_{u}^{2}{Y_{\nu}^{T}  Y_\nu}  + M_{\tilde{L}}^2,
 \\&&m_{{\sigma}_{l}{\sigma}_{r}} = \frac{1}{\sqrt{2} } v_uT_\nu -  v_u v_{\bar{\eta}} {Y_X  Y_\nu}
  - \frac{1}{2}v_d ({\lambda}_{H}v_S  + \sqrt{2} \mu )Y_\nu,\\&&
m_{{\sigma}_{r}{\sigma}_{r}}= \frac{1}{8} \Big((g_{Y X} g_{X}+g_{X}^{2})(v_{d}^{2}- v_{u}^{2})
+2g_{X}^{2}(v_{\eta}^{2}- v_{\bar{\eta}}^{2})\Big)- v_{\eta} v_S Y_X {\lambda}_{C}\nonumber \\&&\hspace{1.8cm}
+M_{\tilde{\nu}}^2 + \frac{1}{2} v_{u}^{2}|Y_\nu|^2+  v_{\bar{\eta}} (2 v_{\bar{\eta}}Y_X  Y_X  - \sqrt{2} T_X).
\end{eqnarray}
 One can find other mass matrixes in the Appendix.\ref{A1} and Ref.\cite{UU1}.

Here, we show some needed couplings in this model.
The charginos interact with  down-quarks and up-squarks
\begin{eqnarray}
\mathcal{L}_{\chi^{-}\tilde{u}d}=\bar{\chi}_i^{-}
\Big((V_{i2}^{*}Y_u^jZ^U_{k,3+j}-g_2V_{i1}^{*}Z_{kj}^U
)P_L+Y_d^{j*}Z_{k,b}^UU_{i2}P_R\Big)d_{j}\tilde{u}_{k}^*.
\end{eqnarray}
We also deduce the vertexes of $Z-\tilde{e}_i-\tilde{e}^{*}_j$
\begin{eqnarray}
\mathcal{L}_{Z\tilde{e}\tilde{e}^{*}}=\frac{1}{2}\tilde{e}^{*}_j
\Big[(g_2\cos\theta_W\cos\theta_W^\prime-g_1\cos\theta_W^\prime\sin\theta_W
+g_{YX}\sin\theta_W^\prime)\sum_{a=1}^3Z_{i,a}^{E,*}Z_{j,a}^E\nonumber\\
+\Big((2g_{YX}+g_X)\sin\theta_W^\prime-2g_1\cos\theta_W^\prime\sin\theta_W\Big)
\sum_{a=1}^3Z_{i,3+a}^{E,*}Z_{j,3+a}^E\Big](p^\mu_{i}-p^\mu_j)\tilde{e}_iZ_{\mu}.
\end{eqnarray}

To save space in the text, the remaining vertexes can be found in Refs.\cite{UU2,UU3,UU4,UU5}.

\section {formulation}
In this section, we study the neutrino mixing, the amplitudes and branching ratios of $l_j\rightarrow{l_i\gamma}$, $\mu\rightarrow{eee}$ and $\tau\rightarrow{Pl}$.

\subsection{Neutrino Mixing}
In this subsection, we use the "top-down" method\cite{NEW1} to derive the  formulas of neutrino mass and mixing angle from the effective neutrino mass matrix.
The seesaw mechanism has an effect on obtaining the tiny neutrino masses.
In this model, the left-handed neutrino(3 generation) and right-handed neutrino(3 generation)
mix together and form a $6\times6$ neutrino mass matrix, whose concrete form in the base $(\nu_L,\bar{\nu}_R)$ is
\begin{eqnarray}
M_{\nu}=
\left({\begin{array}{*{20}{c}}
0 & \frac{\upsilon_u}{\sqrt{2}}(Y_\nu^T)^{IJ}  \\
\frac{\upsilon_u}{\sqrt{2}}(Y_\nu)^{IJ} & \sqrt{2}\upsilon_{\bar{\eta}}(Y_X)^{IJ}  \\
\end{array}}
\right),~~~~~~~ {\rm with}~~~ I,J=1,2,3.
\end{eqnarray}
The effective light neutrino mass matrix is in general given as $m_{eff}=-mM^{-1}m^{T}$,
with
\begin{eqnarray}
&&m=\frac{1}{\sqrt{2}}v_uY^T_\nu,~~~~~M=\sqrt{2}v_{\overline{\eta}}Y_X,\nonumber\\
&&Y_{\nu}=\left(
\begin{array}{ccc}
(Y_{\nu})_{11}&(Y_{\nu})_{12}&(Y_{\nu})_{13}\\
(Y_{\nu})_{12}&(Y_{\nu})_{22}&(Y_{\nu})_{23}\\
(Y_{\nu})_{13}&(Y_{\nu})_{23}&(Y_{\nu})_{33}\\
\end{array}
\right).
\end{eqnarray}

For the effective neutrino mass matrix $m_{eff}$, we obtain the
Hermitian matrix
\begin{eqnarray}
\mathcal{H}=m_{eff}^{\dag}m_{eff}.
\end{eqnarray}
 Here we consider the spectrum with normal ordering(NO), the specific derivation formulas can be found in the Appendix \ref{A2}.

The constraints from neutrino experiment data are \cite{sun10}
\begin{eqnarray}
&&\sin^2(\theta_{12})=0.307^{+0.013}_{-0.012},\nonumber\\
&&\sin^2(\theta_{23})=0.546\pm0.021,\nonumber\\
&&\sin^2(\theta_{13})=0.022\pm0.0007,\nonumber\\
&&\Delta{m^2_\odot}=(7.53\pm0.18)\times10^{-5}~{\rm eV}^2,\nonumber\\
&&|\Delta{m_{A}^2}|=(2.453\pm0.033)\times10^{-3}~{\rm eV}^2.
{\label{E6}}
\end{eqnarray}

\subsection{$l_j\rightarrow{l_i\gamma}$}
In this subsection, we study the lepton flavor violating decays of the $l_j\rightarrow{l_i\gamma}~(j=\tau,\mu,~i=\mu,e$ and $i\neq j)$ processes under the $U(1)_X$SSM model. The Feynman diagrams are shown in Fig.1.
If the external lepton is on shell, the amplitude of $l_j\rightarrow{l_i\gamma}$ is
\begin{eqnarray}
\mathcal{M}=e\varepsilon^\mu\overline{u}_i(p+q)[q^2\gamma_\mu(C^L_1P_L+C^R_1P_R)+m_{l_j}i\sigma_{\mu\nu}q^\nu(C^L_2P_L+C^R_2P_R)]u_j(p),{\label{NI}}
\end{eqnarray}
 where $p$ is the injecting lepton momentum, $q$ is the photon momentum, and $m_{l_j}$ is the mass of the $j$th generation charged lepton. $u_i(p)$ and $v_i(p)$ are the wave functions for the external leptons. The final Wilson coefficients $C^L_1,~C^R_1,~C^L_2,~C^R_2$ are obtained from the sum of these diagrams' amplitudes.
\begin{figure}[h]
\setlength{\unitlength}{5.0mm}
\centering
\includegraphics[width=5.0in]{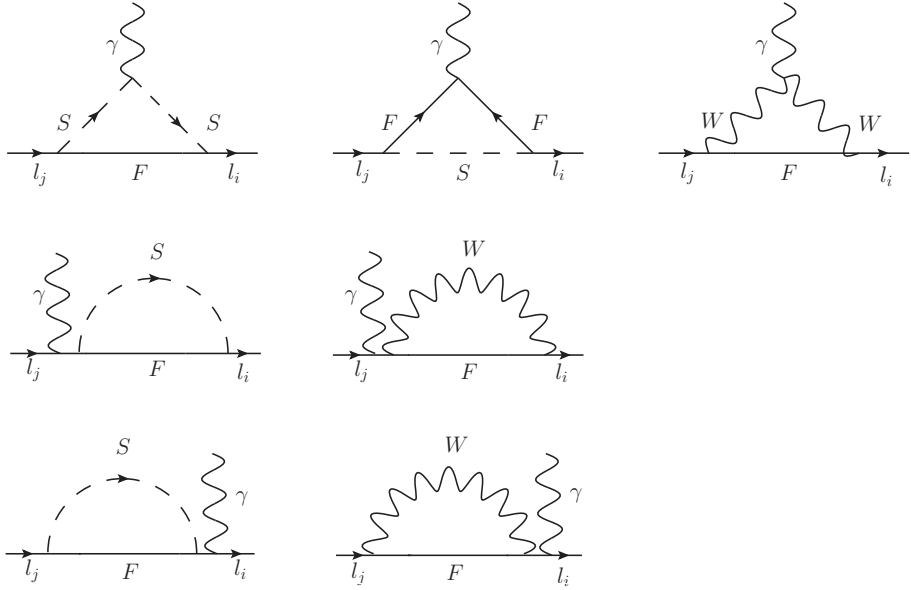}
\caption{ The one-loop diagrams for $l_j\rightarrow{l_i\gamma}$, with $F$ representing Dirac (Majorana) particles.}\label{N1}
\end{figure}

For the specific contribution form of the self-energy diagrams in Fig.\ref{N1}, please refer to \ref{A3}(the subsection D of section III).
 The contributions from the virtual neutral fermion diagrams, the virtual charged fermion diagrams and the virtual W diagrams are expressed by $C^{L,R}_{\alpha}(n),~C^{L,R}_{\alpha}(c),~C^{L,R}_{\alpha}(W)~(\alpha=1,2)$ respectively. The specific analytical results of $C^{L,R}_{\alpha}(n),~C^{L,R}_{\alpha}(c)$ and $C^{L,R}_{\alpha}(W)$ are shown in Ref.\cite{NEW2}.
 Finally, we get the final Wilson coefficient and decay width of $l_j\rightarrow{l_i\gamma}$,
\begin{eqnarray}
&&C^{L,R}_{\alpha}=C^{L,R}_{\alpha}(n)+C^{L,R}_{\alpha}(c)+C^{L,R}_{\alpha}(W),~~\alpha=1,2,\nonumber\\
&&\Gamma(l_j\rightarrow{l_i\gamma})=\frac{e^2}{16\pi}m^5_{l_j}(|C^L_2|^2+|C^R_2|^2).
\end{eqnarray}
The branching ratio of $l_j\rightarrow{l_i\gamma}$ is
\begin{eqnarray}
Br(l_j\rightarrow{l_i\gamma})=\Gamma(l_j\rightarrow{l_i\gamma})/\Gamma_{l_j}.
\end{eqnarray}

\subsection{$\mu\rightarrow{eee}$}
\begin{figure}[h]
\setlength{\unitlength}{5.0mm}
\centering
\includegraphics[width=5.0in]{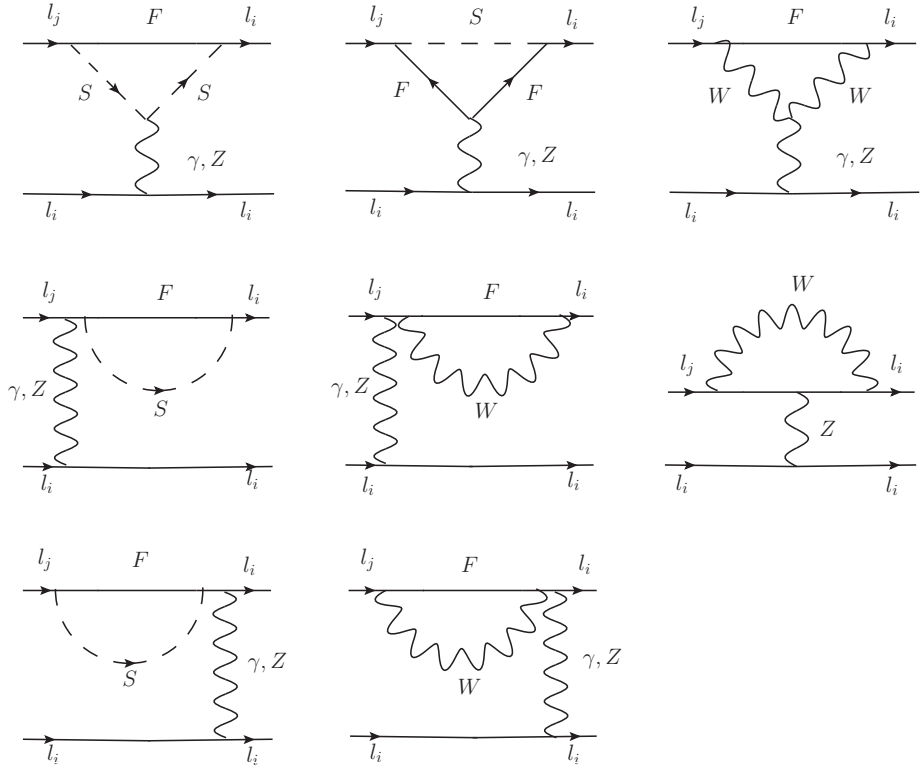}
\caption{ The penguin-type and self-energy-type diagrams for $\mu\rightarrow{eee}$, with $F$ representing Dirac (Majorana) particles.}{\label{N2}}
\end{figure}

The effective Lagrangian of $\mu\rightarrow{eee}$ process is dominated by penguin-type diagrams($\gamma$-penguin and Z-penguin) and box-type diagrams. For the specific contribution of the self-energy diagrams, one can refer to \ref{A3}.
Let's first discuss the contribution of the penguin-type diagrams in Fig.\ref{N2}.

For simplicity, we use Eq.(\ref{NI}) to get the contribution of $\gamma$-penguin diagrams,
\begin{eqnarray}
&&T_{\gamma-p}=\overline{u}_i(p_1)[q^2\gamma_\mu(C^L_1P_L+C^R_1P_R)+m_{l_j}i\sigma_{\mu\nu}q^\nu(C^L_2P_L+C^R_2P_R)]u_j(p)\nonumber\\
&&~~~~~~~~\times\frac{e^2}{q^2}\overline{u}_i(p_2)\gamma^{\mu}v_i(p_3)-(p_1\leftrightarrow{p_2}).
\end{eqnarray}
 The contribution from $Z$-penguin diagrams is in the following:
\begin{eqnarray}
&&T_{Z-p}=\frac{e^2}{m^2_Z}\overline{u}_i(p_1)\gamma_\mu(N_LP_L+N_RP_R)u_j(p)\overline{u}_i(p_2)\gamma^\mu(H^{Zl_i\overline{l}_i}_LP_L+H^{Zl_i\overline{l}_i}_RP_R)v_i(p_3)\nonumber\\
&&~~~~~~~~~-(p_1\leftrightarrow{p_2}),\nonumber\\
&&N_{L,R}=N_{L,R}(S)+N_{L,R}(W).
\end{eqnarray}
$N_{L,R}(S)$ and $N_{L,R}(W)$ represent the effective couplings.
\begin{figure}[h]
\setlength{\unitlength}{5.0mm}
\centering
\includegraphics[width=5.0in]{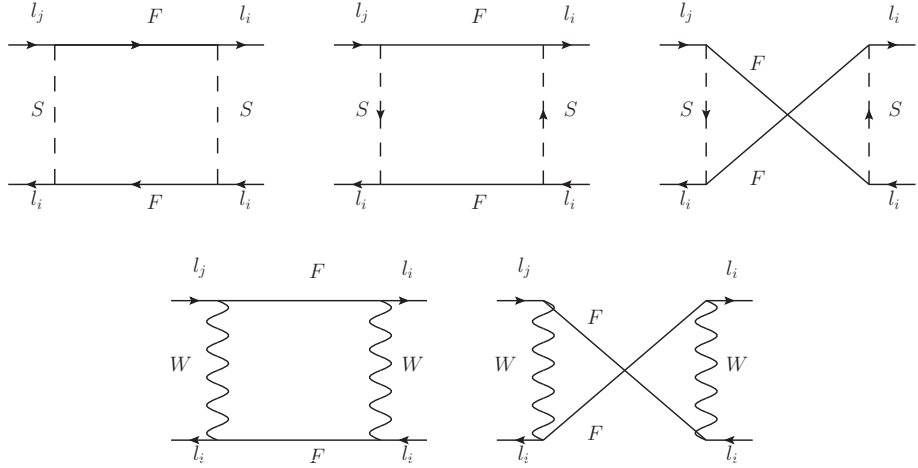}
\caption{ The box-type diagrams for $\mu\rightarrow{eee}$, with $F$ representing Dirac (Majorana) particles.}\label{N3}
\end{figure}

Next, the contribution of the box-type diagrams shown in Fig.\ref{N3} is
\begin{eqnarray}
&&T_{box}=\{B^L_1e^2\overline{u}_i(p_1)\gamma_{\mu}P_Lu_j(p)\overline{u}_i(p_2)\gamma^{\mu}P_L{\nu}_i(p_3)+(L\leftrightarrow{R})\}\nonumber\\
&&+\{B^L_2e^2[\overline{u}_i(p_1)\gamma_{\mu}P_Lu_j(p)\overline{u}_i(p_2)\gamma^{\mu}P_R{\nu}_i(p_3)-(p_1\leftrightarrow{p_2}]+(L\leftrightarrow{R})\}\nonumber\\
&&+\{B^L_2e^3[\overline{u}_i(p_1)P_Lu_j(p)\overline{u}_i(p_2)P_L{\nu}_i(p_3)-(p_1\leftrightarrow{p_2}]+(L\leftrightarrow{R})\}\nonumber\\
&&+\{B^L_2e^4[\overline{u}_i(p_1)\sigma_{\mu\nu}P_Lu_j(p)\overline{u}_i(p_2)\sigma_{\mu\nu}P_L{\nu}_i(p_3)-(p_1\leftrightarrow{p_2}]+(L\leftrightarrow{R})\}\nonumber\\
&&+\{B^L_2e^5[\overline{u}_i(p_1)P_Lu_j(p)\overline{u}_i(p_2)P_R{\nu}_i(p_3)-(p_1\leftrightarrow{p_2}]+(L\leftrightarrow{R})\}.
\end{eqnarray}

 From the box-type diagrams, we obtain the virtual chargino contributions to the effective couplings $B^{L,R}_\beta(c)~(\beta=1\dots5)$.
 The contributions of the neutralino-slepton and W-neutrino to the effective couplings are expressed by $B^{L,R}_\beta(n)$
and $B^{L,R}_\beta(W)~(\beta=1\dots5)$.

To sum up, $B^{L,R}_\beta$ are expressed as
\begin{eqnarray}
B^{L,R}_\beta=B^{L,R}_\beta(n)+B^{L,R}_\beta(c)+B^{L,R}_\beta(W),~~~~~(\beta=1\dots5).
\end{eqnarray}
Because the formulas are cumbersome, $C^{L,R}_\alpha~(\alpha=1,2)$, $N_{L,R}$, $B^{L,R}_\beta~(\beta=1\dots5)$  and the decay width of $\mu\rightarrow{eee}$ can be found in Ref.\cite{NEW2}.
 The branching ratio of $\mu\rightarrow{eee}$ is given by
\begin{eqnarray}
Br(\mu\rightarrow{eee})=\frac{\Gamma(\mu\rightarrow{eee})}{\Gamma_\mu}.
\end{eqnarray}
Here $\Gamma_\mu\simeq3.004\times10^{-19}$.

\subsection{$\tau\rightarrow{Pl}$}{\label{A3}}
\begin{figure}[h]
\setlength{\unitlength}{5.0mm}
\centering
\includegraphics[width=5.0in]{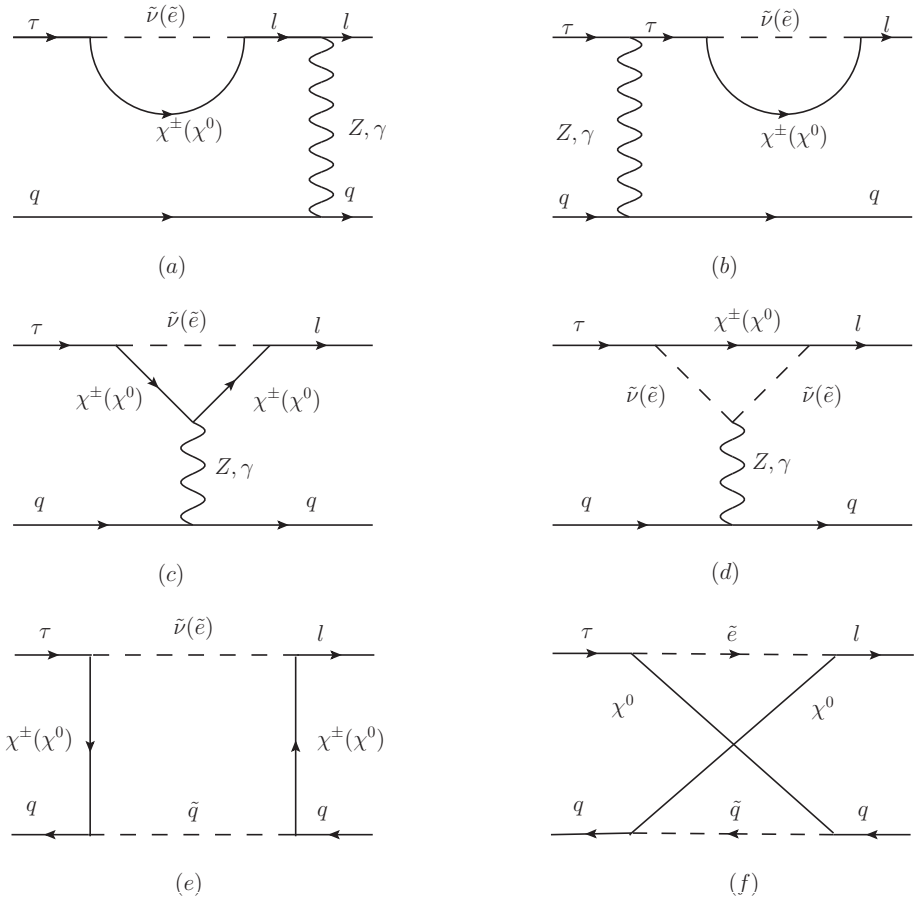}
\caption{The Feynman diagrams contributing to $\tau\rightarrow{Pl}$ in the $U(1)_X$SSM. There are three types of diagrams: self-energy diagrams (a) and (b), penguin diagrams (c) and (d), box diagrams (e) and (f).}{\label{1}}
\end{figure}
At the quark level, the effective Lagrangion for $\tau\rightarrow{Pl}$ can be written as\cite{sun11}
\begin{eqnarray}
&&\mathcal{L}_{\tau\rightarrow{Pl}}=X_{LL}^I(\overline{l}_\alpha{P_L}\tau)(\overline{d}P_Ld)+Y_{LL}^I(\overline{l}_\alpha{P_L}\tau)(\overline{u}{P_L}u)\nonumber\\
&&+X_{LR}^I(\overline{l}_\alpha{P_L}\tau)(\overline{d}{P_R}d)+Y_{LR}^I(\overline{l}_\alpha{P_L}\tau)(\overline{u}{P_R}u)\nonumber\\
&&+X_{LL}^J(\overline{l}_\alpha\gamma_\mu{P_L}\tau)(\overline{d}\gamma^\mu{P_L}d)+Y_{LL}^J(\overline{l}_\alpha\gamma_\mu{P_L}\tau)(\overline{u}\gamma^\mu{P_L}u)\nonumber\\
&&+X_{LR}^J(\overline{l}_\alpha\gamma_\mu{P_L}\tau)(\overline{d}\gamma^\mu{P_R}d)+Y_{LR}^J(\overline{l}_\alpha\gamma_\mu{P_L}\tau)(\overline{u}\gamma^\mu{P_R}u)+(L\leftrightarrow{R})+h.c.,
\end{eqnarray}
where the index $\alpha(=1,2)$ denotes the generation of the emitted lepton and $l_1(l_2)=e(\mu)$.

We show the studied Feynman diagrams contributing to $\tau\rightarrow{Pl}$ in the Fig.\ref {1}.
Taking Fig.\ref {1}(c) as an example, the effective amplitude can be written as
\begin{eqnarray}
&&\mathcal{M}_{(c)}=\sum_{\tilde{\nu}=\tilde{\nu}^I\!,\;\tilde{\nu}^R}\Big[
-\Big(\frac{1}{2m_Z^2}G_3(x_{\tilde{\nu}},x_{\chi^\pm})
A_R^{\tilde{\nu}\bar{l}_i\chi^\pm}B_R^{\bar{\chi}^\pm{Z^0}\chi^\pm}C_L^{\tilde{\nu}\bar{\chi}^\pm\tau}D_L^{\bar{q}Z^0q}
\nonumber\\&&\hspace{1.5cm}+\frac{x_f}{m_Z^2}I_3(x_{\tilde{\nu}},x_{\chi^\pm})A_R^{\tilde{\nu}\bar{l}_i\chi^\pm}
B_L^{\bar{\chi}^\pm{Z^0}\chi^\pm}C_L^{\tilde{\nu}\bar{\chi}^\pm\tau}D_L^{\bar{q}Z^0q}\Big)
(\bar{l}_i\gamma^uP_L\tau)(\bar{q}\gamma_uP_Lq)\nonumber\\
&&\hspace{1.5cm}-\Big(\frac{1}{2m_Z^2}G_3(x_{\tilde{\nu}},x_{\chi^\pm})A_L^{\tilde{\nu}\bar{l}_i\chi^\pm}
B_L^{\bar{\chi}^\pm{Z^0}\chi^\pm}C_R^{\tilde{\nu}\bar{\chi}^\pm\tau}D_L^{\bar{q}Z^0q}\nonumber\\&&\hspace{1.5cm}
+\frac{x_f}{m_Z^2}I_3(x_{\tilde{\nu}},x_{\chi^\pm})A_L^{\tilde{\nu}\bar{l}_i
\chi^\pm}B_R^{\bar{\chi}^\pm{Z^0}\chi^\pm}C_R^{\tilde{\nu}\bar{\chi}^\pm\tau}D_L^{\bar{q}Z^0q}\Big)
(\bar{l}_i\gamma^uP_R\tau)(\bar{q}\gamma_uP_Lq)\nonumber\\
&&\hspace{1.5cm}-\Big(\frac{1}{2m_Z^2}G_3(x_{\tilde{\nu}},x_{\chi^\pm})A_R^{\tilde{\nu}\bar{l}_i
\chi^\pm}B_R^{\bar{\chi}^\pm{Z^0}\chi^\pm}C_L^{\tilde{\nu}\bar{\chi}^\pm\tau}D_R^{\bar{q}Z^0q}
\nonumber\\&&\hspace{1.5cm}+\frac{x_f}{m_Z^2}I_3(x_{\tilde{\nu}},x_{\chi^\pm})A_R^{\tilde{\nu}\bar{l}_i
\chi^\pm}B_L^{\bar{\chi}^\pm{Z^0}\chi^\pm}C_L^{\tilde{\nu}\bar{\chi}^\pm\tau}D_R^{\bar{q}Z^0q}\Big)
(\bar{l}_i\gamma^uP_L\tau)(\bar{q}\gamma_uP_Rq)\nonumber\\
&&\hspace{1.5cm}-\Big(\frac{1}{2m_Z^2}G_3(x_{\tilde{\nu}},x_{\chi^\pm})A_L^{\tilde{\nu}\bar{l}_i\chi^\pm}
B_L^{\bar{\chi}^\pm{Z^0}\chi^\pm}C_R^{\tilde{\nu}\bar{\chi}^\pm\tau}D_R^{\bar{q}Z^0q}
\nonumber\\&&\hspace{1.5cm}+\frac{x_f}{m_Z^2}I_3(x_{\tilde{\nu}},x_{\chi^\pm})A_L^{\tilde{\nu}\bar{l}_i
\chi^\pm}B_R^{\bar{\chi}^\pm{Z^0}\chi^\pm}C_R^{\tilde{\nu}\bar{\chi}^\pm\tau}D_R^{\bar{q}Z^0q}\Big)(\bar{l}_i\gamma^uP_R\tau)(\bar{q}\gamma_uP_Rq)\Big],
\end{eqnarray}
where $x_i=m_i^2/m_W^2$. The functions $G_3(x_1,x_2)$ and $I_3(x_1,x_2)$ are
\begin{eqnarray}
&&G_3(x_1,x_2)=\frac{1}{16\pi^2}[1+\ln x_\mu+\frac{{x_2}^2\ln{x_2}-{x_1}^2\ln{x_1}}{{(x_2-x_1)}^2}
+\frac{x_2+2x_2\ln{x_2}}{x_1-x_2}],\nonumber\\&&
I_3(x_1,x_2)=\frac{1}{16\pi^2}[\frac{1+\ln{x_2}}{x_2-x_1}+\frac{x_1\ln{x_1}-x_2\ln{x_2}}{{(x_2-x_1)}^2}].
\end{eqnarray}
$A,\,B,\,C,\,D$ are coupling constants of the corresponding vertexes, whose concrete forms can be found in Refs.\cite{UU2,UU3,UU4}.

The coefficients $c_P^{I,J}$ and $d_P^{I,J}$ are liner combinations of the Wilson coefficients in the interaction Lagrangion for $\tau\rightarrow{Pl}$\cite{sun9}
\begin{eqnarray}
&&c_P^I=\frac{f_\pi}{2}[\frac{D_L^d(P)}{m_d}(X_{LL}^I+X_{RL}^I)+\frac{D_L^u(P)}{m_u}(Y_{LL}^I+Y_{RL}^I)\nonumber\\&&
\hspace{1.0cm}+\frac{D_R^d(P)}{m_d}(X_{LR}^I+X_{RR}^I)+\frac{D_R^u(P)}{m_u}(Y_{LR}^I+Y_{RR}^I)],\nonumber\\
&&d_P^I=\frac{f_\pi}{2}[\frac{D_L^d(P)}{m_d}(X_{RL}^I-X_{LL}^I)+\frac{D_L^u(P)}{m_u}(Y_{RL}^I-Y_{LL}^I)
\nonumber\\&&\hspace{1.0cm}+\frac{D_R^d(P)}{m_d}(X_{RR}^I-X_{LR}^I)+\frac{D_R^u(P)}{m_u}(Y_{RR}^I-Y_{LR}^I)],\nonumber\\
&&c_P^J=\frac{f_\pi}{4}C(P)(m_\pi-m_l)[X_{LR}^J-X_{LL}^J-X_{RL}^J+X_{RR}^J+Y_{LL}^J-Y_{LR}^J+Y_{RL}^J-Y_{RR}^J],\nonumber\\
&&d_P^J=\frac{f_\pi}{4}C(P)(m_\pi+m_l)[X_{LR}^J-X_{LL}^J+X_{RL}^J-X_{RR}^J+Y_{LL}^J-Y_{LR}^J-Y_{RL}^J+Y_{RR}^J].
\end{eqnarray}

The expressions for coefficients $C(P),D_L^{d,u}(P)$ are listed in Table \ref{II}. Here, $m_\pi$ and $m_K$ denote the masses of neutral pion and Kaon, and $\theta_\eta$ denotes the $\eta\rightarrow\eta^\prime$ mixing angle. In addition, $D_R^{d,u}(P)=-(D_L^{d,u}(P))^*$ and $f_\pi$ is the pion decay constant.\\
\begin{table}
\caption{ Coefficients for each pseudoscalar meson $P$\cite{sun12}}
\begin{tabular}{|c|c|c|c|}
\hline
~~~~ &$C(P)$ & $D^d_L(P)$ & $D^u_L(P)$  \\
\hline
$P=\pi$ & 1 & $ -\frac{m^2_{\pi}}{4}$ & $\frac{m^2_{\pi}}{4}$  \\
\hline
$P=\eta$ & $\frac{1}{\sqrt{6}}(\sin \theta_\eta+\sqrt{2}\cos\theta_\eta)$ & $\frac{1}{4\sqrt{3}}[(3m^2_\pi-4m^2_K)\cos\theta_\eta-2\sqrt{2}m^2_K\sin\theta_\eta]$ &$\frac{1}{4\sqrt{3}}m^2_\pi(\cos\theta_\eta-\sqrt{2}\sin\theta_\eta)$ \\
\hline
$P=\eta^\prime$ & $\frac{1}{\sqrt{6}}(\sqrt{2}\sin \theta_\eta-\cos\theta_\eta)$ & $\frac{1}{4\sqrt{3}}[(3m^2_\pi-4m^2_K)\sin\theta_\eta+2\sqrt{2}m^2_K\cos\theta_\eta]$  &$\frac{1}{4\sqrt{3}}m^2_\pi(\sin\theta_\eta+\sqrt{2}\cos\theta_\eta)$ \\
\hline
\end{tabular}
\label{II}
\end{table}

The averaged squared amplitude can be written as
\begin{eqnarray}
|\mathcal{M}|^2=2m_\tau{m_l}(c_P^Ic_P^{I*}-d_P^Id_P^{I*})+(m_\tau^2+m_l^2-m_P^2)(c_P^Ic_P^{I*}+d_P^Id_P^{I*})\nonumber\\
+2m_\tau{m_l}(c_P^Jc_P^{J*}-d_P^Jd_P^{J*})+(m_\tau^2+m_l^2-m_P^2)(c_P^Jc_P^{J*}+d_P^Jd_P^{J*})\nonumber\\
+2m_\tau{m_l}(c_P^Ic_P^{J*}-d_P^Id_P^{J*})+(m_\tau^2+m_l^2-m_P^2)(c_P^Ic_P^{J*}+d_P^Id_P^{J*})\nonumber\\
+2m_\tau{m_l}(c_P^Jc_P^{I*}-d_P^Jd_P^{I*})+(m_\tau^2+m_l^2-m_P^2)(c_P^Jc_P^{I*}+d_P^Jd_P^{I*}).
\end{eqnarray}
Then the decay width for $\tau\rightarrow{Pl}$ is given by
\begin{eqnarray}
&&\Gamma(\tau\rightarrow{Pl})=\frac{\lambda^{1/2}(m_\tau^2,m_l^2,m_P^2)}{16\pi{m_\tau^3}}\sum_{i,f}|\mathcal{M}|^2.
\end{eqnarray}
The branching ratio of $\tau\rightarrow{Pl}$ is
\begin{eqnarray}
Br(\tau\rightarrow{Pl})=\Gamma(\tau\rightarrow{Pl})/\Gamma_\tau.
\end{eqnarray}
Here $\Gamma_\tau\simeq2.266\times10^{-12}$~{\rm GeV}, and $\Gamma_\tau$ is the total width of $\tau$\cite{UU3}.

\section{numerical results}

In this section, we study the numerical results and consider the constraints from experiments.
 The constraints from lepton flavor violating processes $l_j\rightarrow l_i \gamma$ and $\mu\rightarrow e e e$ are taken into account,
 and we calculate these processes numerically. The constraints from neutrino mass and mixing angles are also researched.
The lightest CP-even Higgs mass $m_{h^0}$=125.1 GeV\cite{su1,su2} is considered. For the mass of the added heavy vector boson $Z^\prime$,
the latest constraint from experiment is $M_{Z^{\prime}}> 5.1$ TeV\cite{xin1}, which is heavier than the former constraints.
 The limits for the masses of
other particles beyond SM are also considered.
The upper bound on the ratio between $M_{Z^\prime}$ and its gauge coupling $g_X$ is
$M_{Z^\prime}/g_X\geq6$ TeV at 99\% C.L.\cite{ZPG1,ZPG2}. Taking into account the constraint from LHC data, $\tan \beta_\eta<1.5$\cite{TanBP}.
 Since $M_{Z^\prime}$ is much larger than $M_{Z}$, the contribution of $Z^\prime$ is very small at the amplitude level, so we won't calculate it here.

Considering the above constraints in the front paragraph, we use the following parameters
\begin{eqnarray}
&&M_S =2.7 ~{\rm TeV},~
T_{\kappa} =1.6~ {\rm TeV}, ~
M_1 =1.2~{\rm TeV},~ M_{BL}=1~{\rm TeV},
~g_{YX}=0.2,\nonumber\\&&
\xi = 17~{\rm TeV},~Y_{X11} =Y_{X22} = Y_{X33} =1,~
g_X=0.33,~\kappa=1,~\lambda_C = -0.08,
\nonumber\\&& M_{BB^\prime}=0.4~{\rm TeV},~
~T_{\lambda_H} = 0.3~{\rm TeV},~
T_{X11} =T_{X22} =T_{X33}= -1~{\rm TeV},~
 l_W = 4~{\rm TeV}^2,
 \nonumber\\&&\lambda_H = 0.1,~
T_{e11} =T_{e22} =T_{e33}= -3~{\rm TeV},~\tan\beta_\eta=0.8,~ B_{\mu} = B_S=1~{\rm TeV}^2,\nonumber\\&&
T_{\lambda_C} = -0.1~{\rm TeV},~\mu=0.5~{\rm TeV},~M^2_{\tilde{E}11}= M^2_{\tilde{E}22}= M^2_{\tilde{E}33}= 3.6~{\rm TeV}^2.
\end{eqnarray}
To simplify the numerical research, we use the relations for the parameters and they vary in the following numerical analysis
\begin{eqnarray}
&&M^2_{\tilde{\nu}11} = M^2_{\tilde{\nu}22}=M^2_{\tilde{\nu}33}=m^2_{\tilde{\nu}},~
M^2_{\tilde{L}11}= M^2_{\tilde{L}22} = M^2_{\tilde{L}33}=m^2_{\tilde{L}},
\nonumber\\&&T_{\nu11}=T_{\nu22}=T_{\nu33}
=T_{\nu},~M^2_{\tilde{L}13}=M^2_{\tilde{L}31}
,~M^2_{\tilde{L}32}=M^2_{\tilde{L}23},~v_S,\nonumber\\
&&M^2_{\tilde{\nu}13}=M^2_{\tilde{\nu}31},~M^2_{\tilde{\nu}23}=M^2_{\tilde{\nu}32},~T_{e13}=T_{e31},~T_{e23}=T_{e32},~\tan\beta.\nonumber\\&&
\end{eqnarray}
Without special statement, the non-diagonal elements of the parameters are supposed as zero.

\subsection{ Neutrino Mixing}

\begin{figure}[ht]
\setlength{\unitlength}{5mm}
\centering
\includegraphics[width=3.0in]{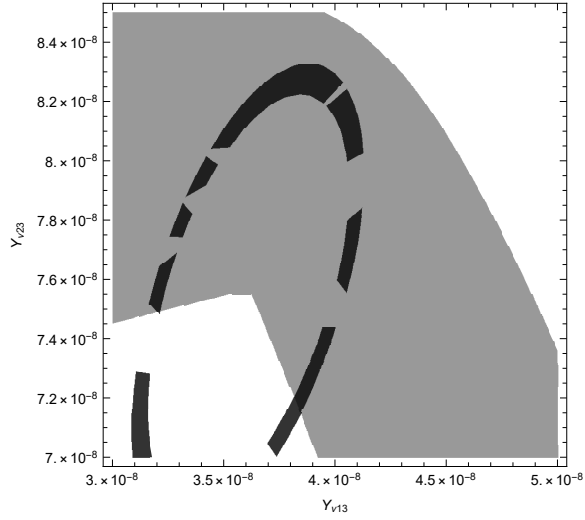}
\vspace{0.2cm}
\caption{Within the range of $3\sigma$, $\Delta{m^2_\odot}$ and $|\Delta{m_{A}^2}|$ are plotted in the plane of $Y_{\nu13}$ versus $Y_{\nu23}$. The black area represents $6.99\times10^{-23}<|\Delta{m_{A}^2}|<8.09\times10^{-23}$ and the gray area represents $2.354\times10^{-21}<\Delta{m^2_\odot}<2.552\times10^{-21}$ .}{\label {N4}}
\end{figure}

 \begin{figure}[ht]
\setlength{\unitlength}{5mm}
\centering
\includegraphics[width=3.0in]{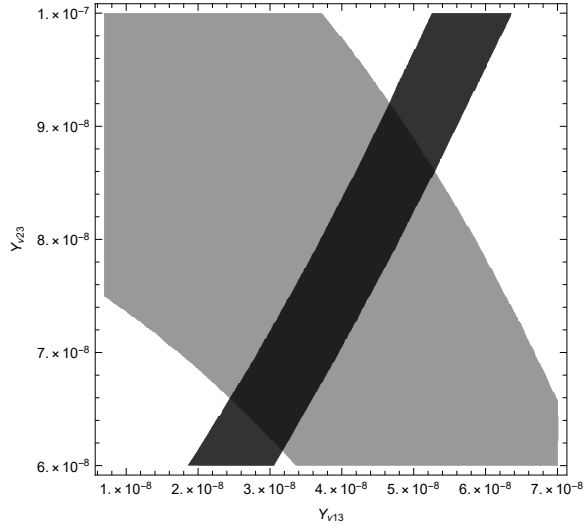}
\vspace{0.2cm}
\caption{Within the range of $3\sigma$, $\sin^2(\theta_{23})$ and $\sin^2(\theta_{13})$ are plotted in the plane of $Y_{\nu13}$ versus $Y_{\nu23}$. The black area represents $0.0199<\sin^2(\theta_{13})<0.0241$ and the gray area represents $0.483<\sin^2(\theta_{23})<0.609$ .}{\label {N5}}
\end{figure}
In this subsection, with $\tan\beta=9$ we
fit the experimental data of neutrino mass variances and mixing angles in Eq.(\ref{E6}) for the normal order condition.
We take the range size of $3\sigma$ and draw the following Figs.\ref{N4}, \ref{N5}.
In Fig.\ref{N4}, $\Delta{m_{A}^2}$ and $\Delta{m^2_\odot}$ are plotted in the plane of $Y_{\nu23}$ versus $Y_{\nu13}$.
Here, the three constraints from the mixing angles are satisfied.
The black area is $6.99\times10^{-23}<|\Delta{m_{A}^2}|<8.09\times10^{-23}$ and the gray area denotes $2.354\times10^{-21}<\Delta{m^2_\odot}<2.552\times10^{-21}$.
We can see that where the two areas overlap is the part meeting the above two mass variances.
Similarly, in Fig.\ref{N5}, $\sin^2(\theta_{13})$ and $\sin^2(\theta_{23})$ are
plotted in the plane of $Y_{\nu23}$ versus $Y_{\nu13}$.
In this condition, the constraints from two mass variances and
the mixing angle $\sin^2(\theta_{12})$ are satisfied.
The black area and gray area represent $0.0199<\sin^2(\theta_{13})<0.0241$ and $0.483<\sin^2(\theta_{23})<0.609$ respectively,
 and the overlapping part is  needed.

Combining the two figures, we can find appropriate parameter space to satisfy the five constraints from neutrino experiment data in Eq.(\ref{E6}).
 We need to take the common part of the area of the two figures. Finally, we find that the range of $Y_{\nu13}$ is about $3.7\times10^{-8}\sim4.2\times10^{-8}$ and the range of $Y_{\nu23}$ is from $7.2\times10^{-8}$ to $8.3\times10^{-8}$.
In the end, we get the values of $Y_\nu$ to fit Eq.(\ref{E6})
\begin{eqnarray}
&&Y_{\nu11}=8.77211\times10^{-6},~~~~~
Y_{\nu22}=8.87025\times10^{-6},\nonumber\\
&&Y_{\nu33}=8.78768\times10^{-6},~~~~~
Y_{\nu13}=4.07916\times10^{-8},\nonumber\\
&&Y_{\nu23}=7.7777\times10^{-8},~~~~~~
Y_{\nu12}=6.0884\times10^{-8}.
\end{eqnarray}
With these parameters, we calculate the three light neutrino masses
\begin{eqnarray}
m_{\nu1}=0.15693~ {\rm eV}, ~~~m_{\nu2}=0.157168~{\rm eV}, ~~~m_{\nu3}=0.164511~{\rm eV}.
\end{eqnarray}

\subsection{ The processes of $l_j\rightarrow{l_i\gamma}$ and $\mu\rightarrow{eee}$ }
We consider the effects of $l_j\rightarrow{l_i\gamma}$ and $\mu\rightarrow{eee}$ on LFV. After studying lepton flavor violating processes of $\mu\rightarrow{e\gamma}$ and $\mu\rightarrow{eee}$, we find that the off-diagonal terms for the soft breaking slepton mass matrices $M^2_{\tilde{L},\tilde{E}}$ and the trilinear coupling matrix $T_e$ are important. For a more intuitive explanation, we define
\begin{eqnarray}
&&M^2_{\tilde{L}}=\left(
\begin{array}{ccc}
1&\delta^{LL}_{12}&\delta^{LL}_{13}\\
\delta^{LL}_{12}&1&\delta^{LL}_{23}\\
\delta^{LL}_{13}&\delta^{LL}_{23}&1\\
\end{array}
\right)M_L^2,\nonumber\\
&&M^2_{\tilde{E}}=\left(
\begin{array}{ccc}
1&\delta^{RR}_{12}&\delta^{RR}_{13}\\
\delta^{RR}_{12}&1&\delta^{RR}_{23}\\
\delta^{RR}_{13}&\delta^{RR}_{23}&1\\
\end{array}
\right)M_E^2,\nonumber\\
&&T_e=\left(
\begin{array}{ccc}
1&\delta^{LR}_{12}&\delta^{LR}_{13}\\
\delta^{LR}_{12}&1&\delta^{LR}_{23}\\
\delta^{LR}_{13}&\delta^{LR}_{23}&1\\
\end{array}
\right)A_e.
\end{eqnarray}
We find that $\mu\rightarrow{e\gamma}$ and $\mu\rightarrow{eee}$ are greatly affected by $\delta^{XX}_{12} (X=L,R)$. So we set $\delta^{XX}_{12}=0$, the limits of  $\mu\rightarrow{e\gamma}$ and $\mu\rightarrow{eee}$ are met.
In addition, for the $\tau\rightarrow\mu\gamma$ and $\tau\rightarrow{e\gamma}$ processes, they have been met in this paper.

\subsection{ The processes of $\tau\rightarrow{P\mu}$}
In order to obtain reasonable numerical results, we need to study some sensitive parameters and important mass matrices. Then, to show the numerical results clearly, we will discuss the processes of
$\tau\rightarrow{Pe}$ and $\tau\rightarrow{P\mu}$ in two subsections.
We draw the relation diagrams and scatter diagrams of $Br(\tau\rightarrow{Pl})$
with different parameters. After analyzing these graphs and the experimental limits of the branching ratios, reasonable parameter spaces are found to explain LFV.

\begin{figure}[ht]
\setlength{\unitlength}{5mm}
\centering
\includegraphics[width=3.0in]{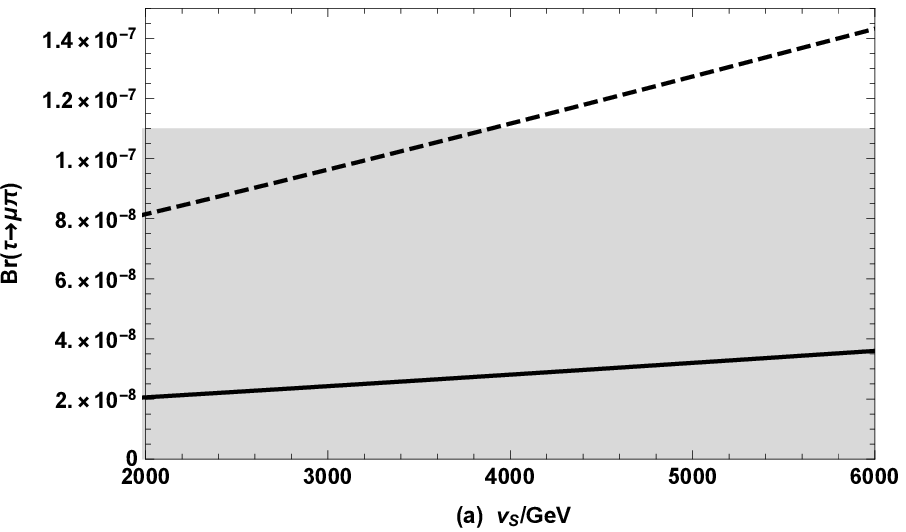}
\vspace{0.2cm}
\setlength{\unitlength}{5mm}
\centering
\includegraphics[width=3.0in]{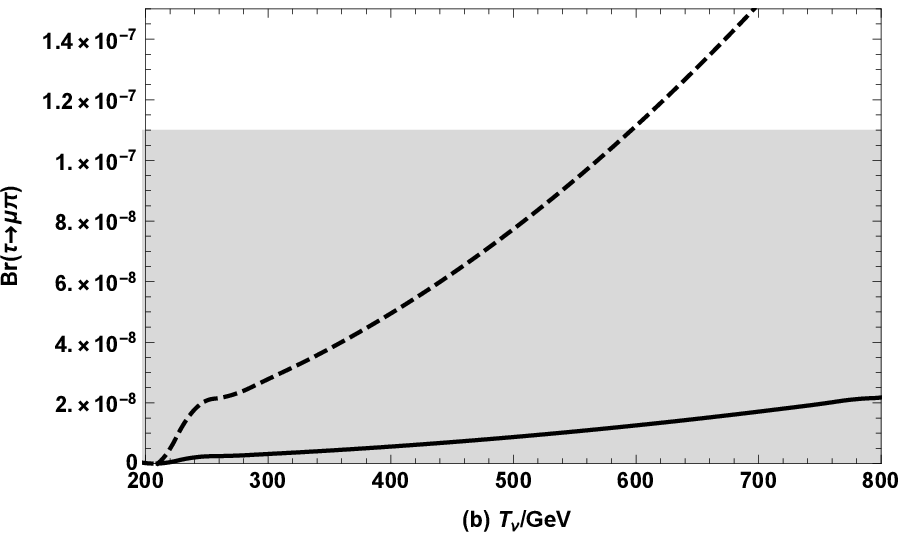}
\vspace{0.2cm}
\setlength{\unitlength}{5mm}
\centering
\includegraphics[width=3.0in]{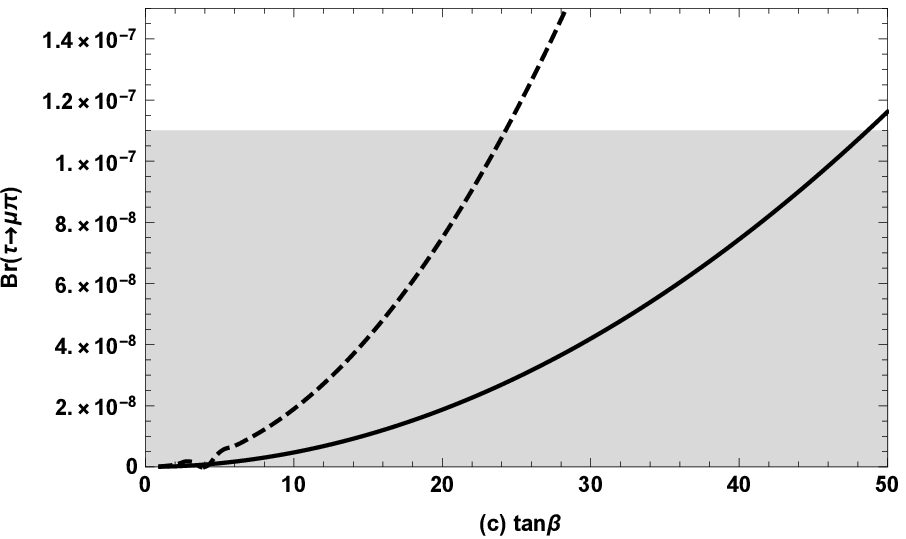}
\caption{$Br(\tau\rightarrow{\mu\pi})$ schematic diagrams affected by different parameters. The gray area is reasonable value range where $Br(\tau\rightarrow\mu\pi)$ is lower than the upper limit.
As $T_{\nu}=500$ GeV, the dashed and solid lines in Fig.\ref {2}(a) correspond to $\tan\beta=40$ and $\tan\beta=20$.
We set $v_S=4300$ GeV in Fig.\ref {2}(b),(c). The dashed line and solid line respectively represent $\tan\beta=30$ and 10 in Fig.\ref {2}(b).
When $T_{\nu}=800$ GeV and 400 GeV, the dashed line and solid line in Fig.\ref {2}(c) are generated.}{\label {2}}
\end{figure}

 With the parameters $M_2=1$ TeV,~$m_{\tilde{\nu}}^2=3\times10^5~{\rm GeV}^2,
 ~m_{\tilde{L}}^2=5\times10^5~\rm TeV^2$,~$M_{\tilde{\nu}13}^2~(M_{\tilde{\nu}23}^2)=100~\rm GeV^2$, we paint $Br(\tau\rightarrow{\mu\pi})$ schematic diagrams affected by different parameters in the Fig.\ref {2}, where the gray area is current limit
  on LFV decay $\tau\rightarrow{\mu\pi}$. Setting $T_{\nu}=500$ GeV, we plot $Br(\tau\rightarrow{\mu\pi})$
  versus $v_S$ in the Fig.\ref {2}(a). The dashed curve corresponds to $\tan\beta=40 $ and
  the solid line corresponds to $\tan\beta=20$. We can clearly see that the two
   lines increase with the increasing $v_S$ in the range of ($2000\sim6000$) GeV. The dashed curve is larger than the solid curve. The solid line part as a whole and the dashed line part of 2000 GeV$<v_S<4000 $ GeV are in the gray area and the dashed line of the rest exceed the gray area.

 Supposing $v_S=4300$ GeV, we show $Br(\tau\rightarrow{\mu\pi})$ varying with $T_{\nu}$ by the solid curve ($\tan\beta=10$) and dashed curve ($\tan\beta=30$) in the Fig.\ref {2}(b). During the $T_\nu$ region ($200\sim 800$) GeV, both the dashed line and the solid line are increasing functions, and the slope of the dashed line is greater than that of the solid line. The dashed line in the $T_\nu$ range of 200 GeV to 600 GeV and solid line in the $T_\nu$ range of 200 GeV to 800 GeV are in the gray area.

 Then we analyze the effect of the parameter $\tan\beta$ on branching ratio of $\tau\rightarrow{\mu\pi}$ and try to find the reasonable ranges. Based on $v_S=4300$ GeV, the numerical results are shown in the Fig.\ref {2}(c) by the dashed curve and solid curve corresponding to $T_{\nu}=800 $ GeV and $T_{\nu}=400 $ GeV respectively. $Br(\tau\rightarrow{\mu\pi})$ varies with $\tan\beta$ in the range from 0.5 to 50. It can be clearly seen that both the solid line and the dashed line have an upward trend. The rising range of the dashed line is greater than that of the solid line.
 The dashed line in $\tan\beta$ range of $0.5\sim 25$ and
 the solid line in $\tan\beta$ range of $0.5\sim 48$ are in the gray area.

$Br(\tau\rightarrow\mu\pi)$ increases as the parameters $v_S,~T_{\nu}$ and $\tan\beta$ increase. In the Fig.\ref {2}(a)(b), the dashed line has a higher slope than the solid line, and the overall direction of the solid line
is closer to the experimental upper limit. As $\tan\beta$ goes up,
$Br(\tau\rightarrow\mu\pi)$ gets closer to the experimental upper limit. The smaller $T_{\nu}$ is, the closer the whole thing is
to satisfy the experimental value in the Fig.\ref {2}(c). All in all, $\tan\beta$ and $T_{\nu}$ are sensitive parameters that have great impacts on $Br(\tau\rightarrow{\mu\pi})$.

\begin{figure}[h]
\setlength{\unitlength}{5mm}
\centering
\includegraphics[width=3.0in]{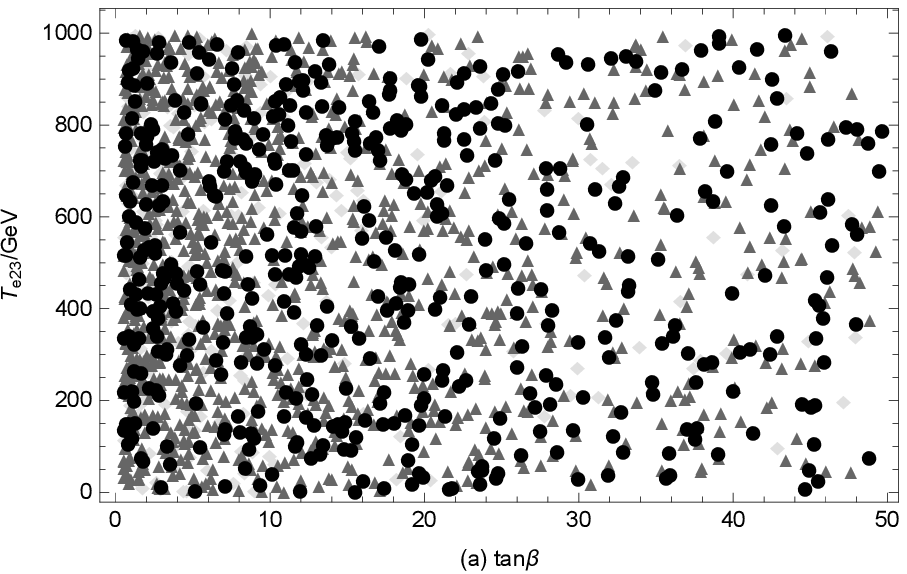}
\vspace{0.2cm}
\setlength{\unitlength}{5mm}
\centering
\includegraphics[width=3.0in]{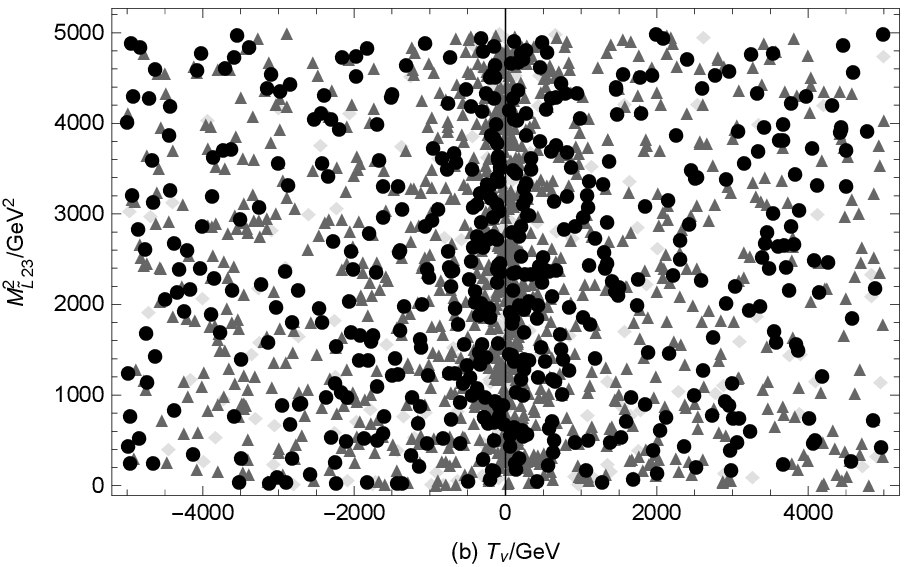}
\vspace{0.2cm}
\setlength{\unitlength}{5mm}
\centering
\includegraphics[width=3.0in]{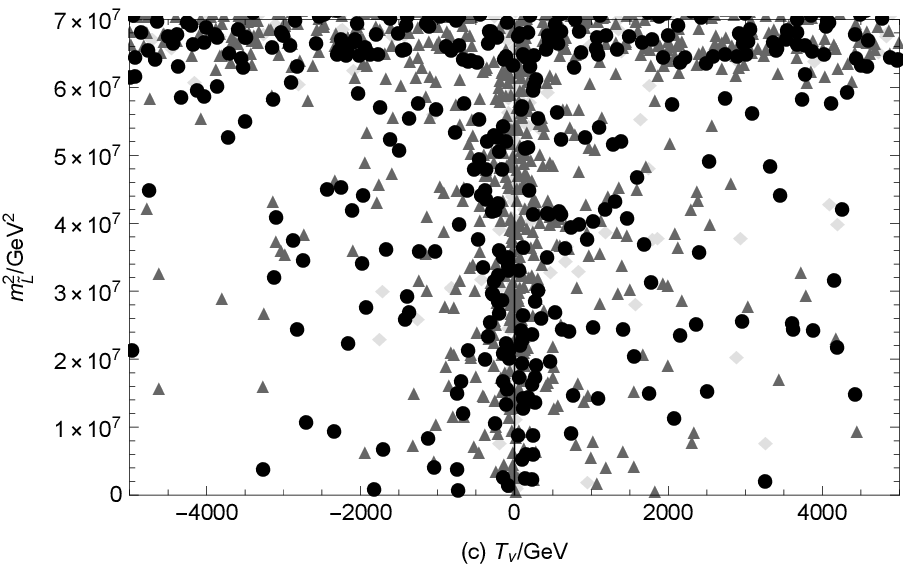}
\vspace{0.2cm}
\setlength{\unitlength}{5mm}
\centering
\includegraphics[width=3.0in]{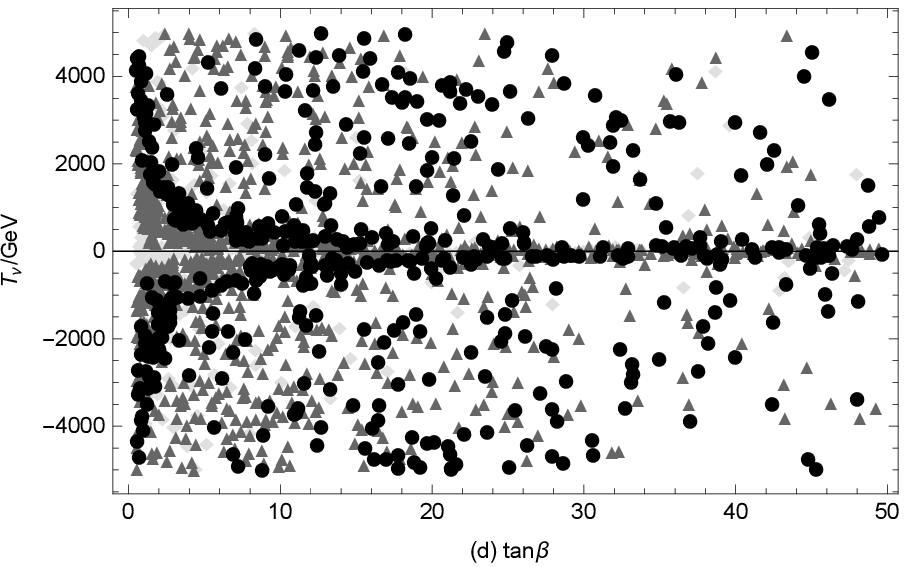}
\caption{Under the premise of current limit on lepton flavor violating decay $\tau\rightarrow\mu\pi$, reasonable parameter space is selected to scatter points, where \textcolor{light-gray} {$\blacklozenge$} mean the value of $Br(\tau\rightarrow\mu\pi)$ less than $0.1\times10^{-8}$, \textcolor{dark-gray}{$\blacktriangle$} mean $Br(\tau\rightarrow\mu\pi)$ in the range of $0.1\times10^{-8}$ to $0.5\times10^{-7}$, $\bullet$ show $0.5\times10^{-7}$ $\leq Br(\tau\rightarrow\mu\pi)<1.1\times10^{-7}$.}
{\label {3}}
\end{figure}

Next, we randomly scan the parameters. Some parameter ranges are set as: $\tan\beta$ from 0.5 to 50, $v_S$ from 2000 GeV to 7000 GeV, $T_{\nu}$ from -5000 GeV to 5000 GeV, and $M_{\tilde{L}23}^2$ ($M_{\tilde{E}23}^2$) from 0  to 5000 $\rm GeV^2$.
In addition, other parameter spaces are represented in tabular form. Fig.\ref {3} is obtained from the parameters shown in the Table \ref {III}. We use \textcolor{light-gray} {$\blacklozenge$} (0$<Br(\tau\rightarrow\mu\pi)<0.1\times10^{-8}$), \textcolor{dark-gray}{$\blacktriangle$} ($0.1\times10^{-8}\leq Br(\tau\rightarrow\mu\pi)<0.5\times10^{-7}$) and $\bullet$ ($0.5\times10^{-7}$ $\leq Br(\tau\rightarrow\mu\pi)<1.1\times10^{-7}$) to
represent the results in different parameter spaces for the process of $\tau\rightarrow{\mu\pi}$.

 The relationship between $\tan\beta$ and $T_{e23}$ is shown in Fig.\ref {3}(a).
All points are mainly concentrated in $0.5<\tan\beta<10$, and with the increase of $\tan\beta$, $Br(\tau\rightarrow{\mu\pi})$ also increases.
Fig.\ref {3}(b) is plotted in the plane of $T_{\nu}$ versus $M_{\tilde{L}23}^2$, and the points mostly concentrate near the $T_{\nu}=0$ coordinate axis.
Here, three types of points are concentrated in -2000 GeV$<T_{\nu}<2000$ GeV.
\textcolor{dark-gray}{$\blacktriangle$} are mainly in -50 GeV$<T_{\nu}<50$ GeV and $\bullet$ are mainly  in -1500 GeV$<T_{\nu}<1500$ GeV.
The effects of $T_{\nu}$ and $m_{\tilde{L}}^2$ on $Br(\tau\rightarrow{\mu\pi})$ are shown in the Fig.\ref {3}(c).
It is obvious that the three types of points are concentrated near the $T_{\nu}=0$ coordinate axis.
Within the range of $6\times10^7 ~{\rm GeV^2}<m_{\tilde{L}}^2<7\times10^7 ~\rm {GeV^2}$, the distributions of \textcolor{light-gray} {$\blacklozenge$} and \textcolor{dark-gray}{$\blacktriangle$} are more concentrated than that of $\bullet$.
Fig.\ref {3}(d) shows the effect of $\tan\beta$ and $T_{\nu}$ on $Br(\tau\rightarrow{\mu\pi})$. On the whole, it is symmetrical about the $T_{\nu}=0$ axis and the distribution of \textcolor{dark-gray}{$\blacktriangle$} is similar to $\bullet$ distribution. They show a downward trend above $T_{\nu}=0$ and gradually approach this line. It looks like an increasing function of the relative distribution below $T_{\nu}=0$. The difference is that \textcolor{dark-gray}{$\blacktriangle$} are closer to $T_{\nu}=0$ than $\bullet$.
\begin{table*}
\caption{Scanning parameters for Fig.{\ref {3}}}\label{III}
\begin{tabular*}{\textwidth}{@{\extracolsep{\fill}}lllll@{}}
\hline
Parameters&Min&Max\\
\hline
$m^2_{\tilde{L}}/\rm GeV^2$&$5\times10^5$&$7\times10^7$\\
$M^2_{\tilde{\nu}13}/\rm GeV^2$&0&$10^4$\\
$M^2_{\tilde{\nu}23}/\rm GeV^2$&0&$10^4$\\
$T_{e13}$/GeV&0&1000\\
$T_{e23}$/GeV&0&1000\\
\hline
\end{tabular*}
\end{table*}

\begin{figure}[h]
\setlength{\unitlength}{5mm}
\centering
\includegraphics[width=2.8in]{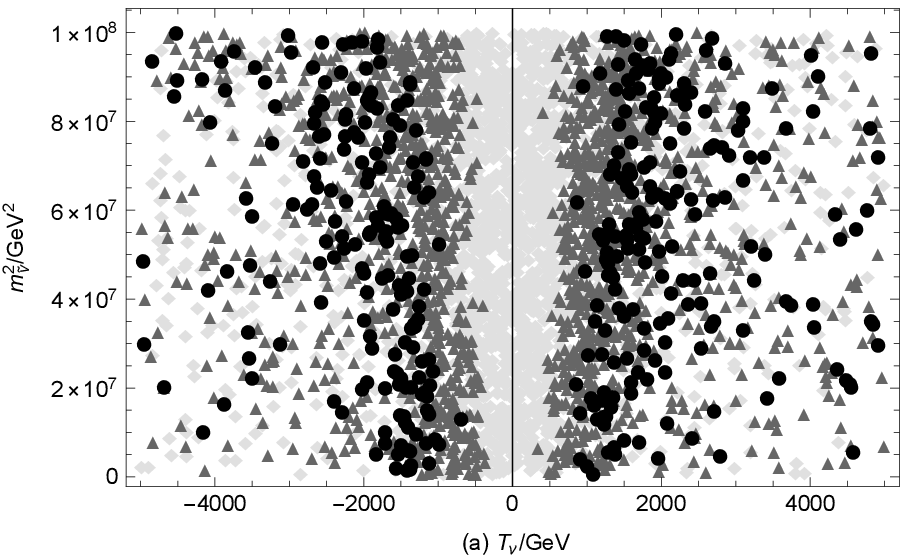}
\vspace{0.2cm}
\setlength{\unitlength}{5mm}
\centering
\includegraphics[width=2.8in]{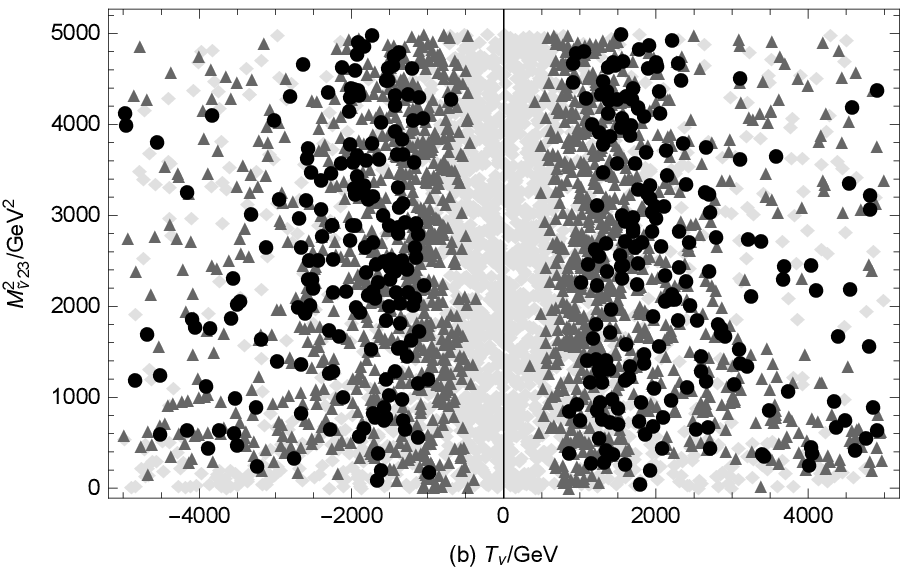}
\vspace{0.2cm}
\setlength{\unitlength}{5mm}
\centering
\includegraphics[width=2.8in]{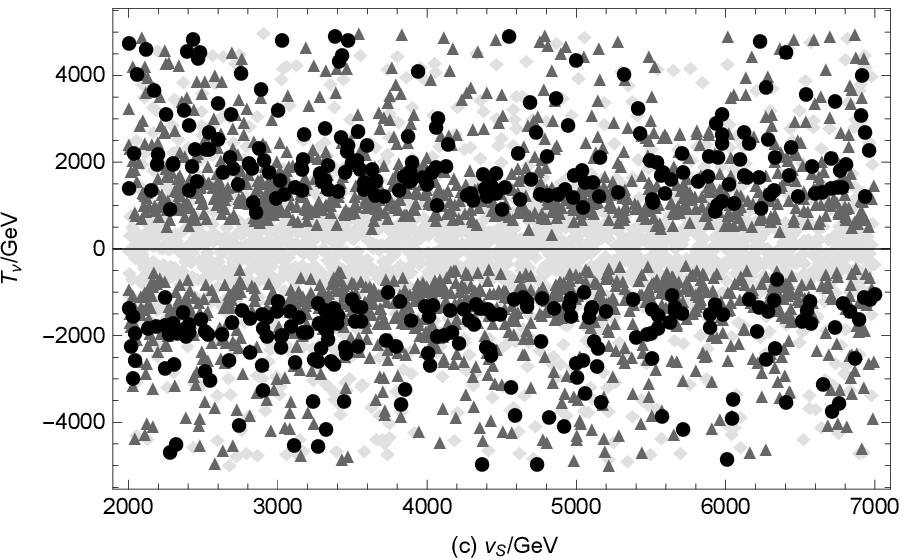}
\caption{Based on the experimental upper limit of $Br(\tau\rightarrow\mu\eta)$, the relationship between parameter space and $Br(\tau\rightarrow\mu\eta)$ is studied.
 As shown in Figs.\ref {4}(a),(b),(c), \textcolor{light-gray} {$\blacklozenge$}, \textcolor{dark-gray}{$\blacktriangle$} and $\bullet$ represent 0$<Br(\tau\rightarrow\mu\eta)<1.0\times10^{-8}$, $1.0\times10^{-8}\leq Br(\tau\rightarrow\mu\eta)<5.0\times10^{-8}$ and $5.0\times10^{-8}\leq Br(\tau\rightarrow\mu\eta)<6.5\times10^{-8}$.}{\label{4}}
\end{figure}

In order to better study LFV and find a reasonable parameter space in the process of
$\tau\rightarrow{\mu\eta}$, we study the effects of parameters $v_S$, $\tan\beta$ and $T_{\nu}$,
and  draw the scatter diagrams of a certain parameter space in Fig.\ref {4}.
We scatter points according to the parameters given in Table \ref {IV} to obtain Figs.\ref {4}(a),(b),(c).
These points are divided into \textcolor{light-gray} {$\blacklozenge$} (0$<Br(\tau\rightarrow\mu\eta)<1.0\times10^{-8}$), \textcolor{dark-gray}{$\blacktriangle$} ($1.0\times10^{-8}$ $\leq Br(\tau\rightarrow\mu\eta)<5.0\times10^{-8}$) and $\bullet$ ($5.0\times10^{-8}$$ \leq Br(\tau\rightarrow\mu\eta)<6.5\times10^{-8}$). In Fig.\ref {4}(a), we can see that $T_{\nu}$ corresponds to $m_{\tilde{\nu}}^2$ and \textcolor{light-gray} {$\blacklozenge$} concentrate in -500 GeV$<T_{\nu}<500$ GeV. \textcolor{dark-gray}{$\blacktriangle$} mainly concentrate in -1000 GeV$<T_{\nu}<-500$ GeV and 500 GeV$<T_{\nu}<1000$ GeV area, and $\bullet$ are scattered between -2500 GeV$<T_{\nu}<-1000$ GeV and 1000 GeV$<T_{\nu}<2500$ GeV. From the perspective of ordinate, the smaller the value of $m_{\tilde{\nu}}^2$, the more blank areas are.
We conclude that the smaller the value of $m_{\tilde{\nu}}^2$, the larger the value of $Br(\tau\rightarrow\mu\eta)$ is. Therefore, in order to meet the experimental limits, we should take the value of $m_{\tilde{\nu}}^2$ as large as possible.

We plot $M_{\tilde{\nu}23}^2$ varying with $T_{\nu}$ in the Fig.\ref {4}(b). The three types of points are concentrated near $T_{\nu}=0$ and $M_{\tilde{\nu}23}^2=0$.  \textcolor{light-gray} {$\blacklozenge$} are mainly in  -500 GeV$<T_{\nu}<500$ GeV and 0$<M_{\tilde{\nu}23}^2<250 \rm GeV^2$. \textcolor{dark-gray}{$\blacktriangle$} are mainly
in -1000 GeV$<T_{\nu}<-500$ GeV, 500 GeV$<T_{\nu}<1000$ GeV and $250{\rm GeV^2}<M_{\tilde{\nu}23}^2<750 \rm GeV^2$. $\bullet$ are mainly in -2250 GeV$<T_{\nu}<-1000$ GeV and 1000 GeV$<T_{\nu}<2250$ GeV. For $M_{\tilde{\nu}23}^2$, \textcolor{light-gray} {$\blacklozenge$}, \textcolor{dark-gray}{$\blacktriangle$} and $\bullet$ are distributed from bottom to top, which once again shows that when $M_{\tilde{\nu}23}^2$ is smaller, $Br(\tau\rightarrow\mu\eta)$ is smaller.

Finally, we analyze the effect from parameters $v_S$ and $T_{\nu}$ in Fig.\ref {4}(c). All points are almost symmetrically distributed about $T_{\nu}=0$ and concentrate in the -2000 GeV$<T_{\nu}<2000$ GeV region.
The farther away from the $T_{\nu}=0$ axis, the more sparse the distribution is. It implies that $v_S$ is a dull parameter,
and affects the results slightly.

\begin{table*}
\caption{Scanning parameters for Fig.{\ref {4}}}
\begin{tabular*}{\textwidth}{@{\extracolsep{\fill}}lllll@{}}
\hline
Parameters&Min&Max\\
\hline
$M_2$/GeV&500&4000\\
$M^2_{\tilde{\nu}13}/\rm GeV^2$&$0$&$5000$\\
$M^2_{\tilde{\nu}23}/\rm GeV^2$&$0$&$5000$\\
$m^2_{\tilde{L}}/\rm GeV^2$&$5\times10^5$&$10^7$\\
$m^2_{\tilde{\nu}}/\rm GeV^2$&$5\times10^5$&$10^8$\\
\hline
\end{tabular*}
\label{IV}
\end{table*}

\begin{figure}[h]
\setlength{\unitlength}{5mm}
\centering
\includegraphics[width=3.0in]{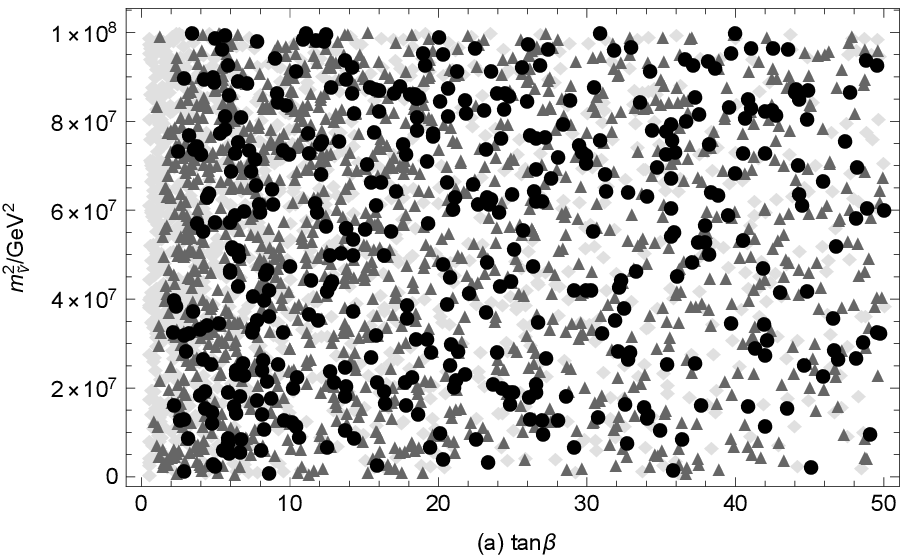}
\vspace{0.2cm}
\setlength{\unitlength}{5mm}
\centering
\includegraphics[width=3.0in]{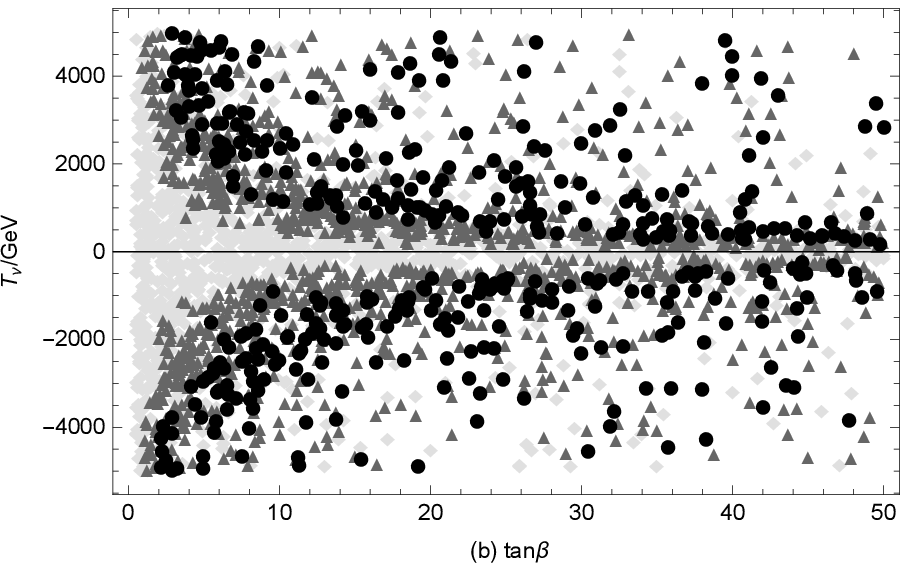}
\caption{ Scatter points in an area smaller than the upper limit of $Br(\tau\rightarrow\mu\eta^\prime)$. \textcolor{light-gray} {$\blacklozenge$} represent 0$<Br(\tau\rightarrow\mu\eta^\prime)<0.1\times10^{-7}$, \textcolor{dark-gray}{$\blacktriangle$} represent $0.1\times10^{-7}\leq Br(\tau\rightarrow\mu\eta^\prime)<0.8\times10^{-7}$ and $0.8\times10^{-7}\leq Br(\tau\rightarrow\mu\eta^\prime)<1.3\times10^{-7}$ are represented by $\bullet$.}{\label {5}}
\end{figure}

For the $\tau\rightarrow{\mu\eta^\prime}$ process, we sprinkle the parameters shown in Table \ref {V}.
In the Fig.\ref {5}, \textcolor{light-gray} {$\blacklozenge$}, \textcolor{dark-gray}{$\blacktriangle$}
and $\bullet$ indicate the ranges of $Br(\tau\rightarrow{\mu\eta^\prime})$ as $(0\sim0.1)\times10^{-7}$,
$(0.1\sim0.8)\times10^{-7}$ and $(0.8\sim1.3)\times10^{-7}$ respectively. Fig.\ref {5}(a) shows that $\tan\beta$ corresponds to $m^2_{\tilde{\nu}}$. The distribution of points is more concentrated in $0.5<\tan\beta<20$. In $0.5<\tan\beta<2$, \textcolor{light-gray} {$\blacklozenge$} are more concentrated, and \textcolor{dark-gray}{$\blacktriangle$} are mainly in $2<\tan\beta<10$. Fig.\ref {5}(b) is shown in the plane of $T_{\nu}$ versus $\tan\beta$, where
the centralized distributions of the three types of points have the same trend. It is almost symmetrical about $T_{\nu}=0$.
In the 0$<T_{\nu}<5000$ GeV region, the subtraction function gradually approaches the $T_{\nu}=0$ axis. It is worth noting that in the $0.5<\tan\beta<20$ part \textcolor{light-gray} {$\blacklozenge$}, \textcolor{dark-gray}{$\blacktriangle$} and $\bullet$ are away from the $T_{\nu}=0$ axis in turn.

\begin{table*}
\caption{Scanning parameters for Fig.{\ref {5}}}\label{V}
\begin{tabular*}{\textwidth}{@{\extracolsep{\fill}}lllll@{}}
\hline
Parameters&Min&Max\\
\hline
$M_2$/GeV&500&4000\\
$M^2_{\tilde{\nu}13}/\rm GeV^2$&$0$&$5000$\\
$M^2_{\tilde{\nu}23}/\rm GeV^2$&$0$&$5000$\\
$m^2_{\tilde{L}}/\rm GeV^2$&$5\times10^5$&$10^7$\\
$m^2_{\tilde{\nu}}/\rm GeV^2$&$5\times10^5$&$10^8$\\
\hline
\end{tabular*}
\end{table*}

\subsection{ The processes of $\tau\rightarrow{Pe}$}
In this subsection, scatter points are in the parameter spaces based on $0.5<\tan\beta<50$,
2000 GeV$<v_S<$ 7500 GeV, -5000 GeV$<T_{\nu}<$ 5000 GeV, 500 GeV$<M_2<$ 4000 GeV and
 $M_{\tilde{\nu}23}^2,~M_{\tilde{L}13}^2,~M_{\tilde{\nu}13}^2,~M_{\tilde{E}13}^2$  varying from 0 to 5000 $\rm GeV^2$.
 Other parameter spaces are represented in tabular form.

\begin{figure}[h]
\setlength{\unitlength}{5mm}
\centering
\includegraphics[width=3.0in]{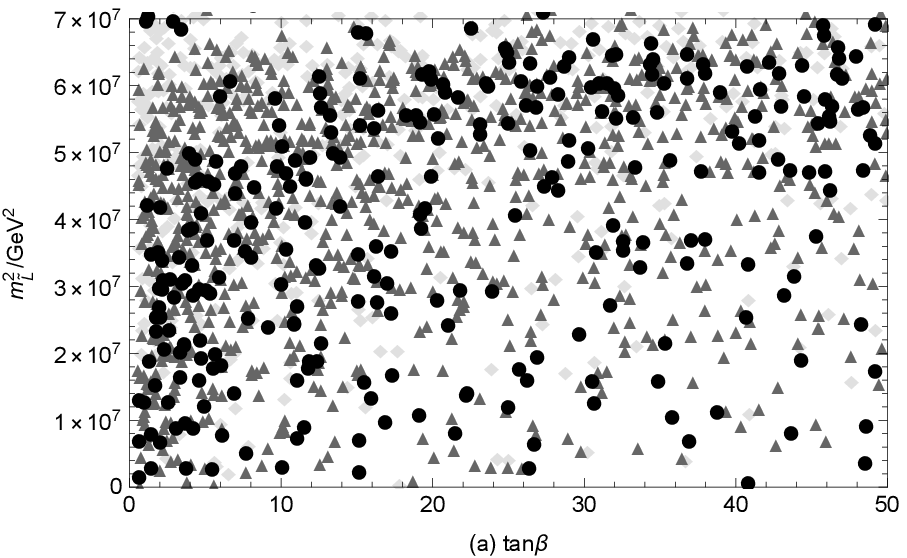}
\vspace{0.2cm}
\setlength{\unitlength}{5mm}
\centering
\includegraphics[width=3.0in]{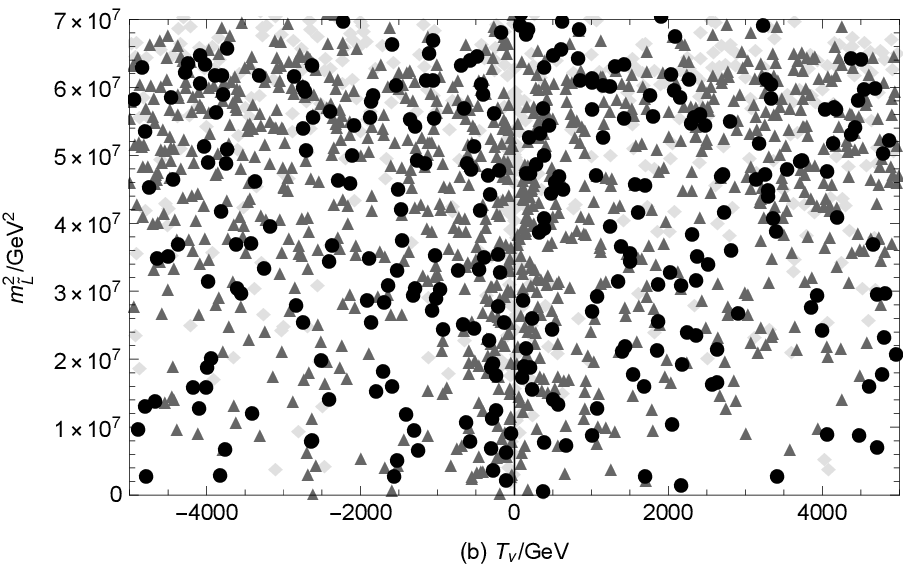}
\vspace{0.2cm}
\setlength{\unitlength}{5mm}
\centering
\includegraphics[width=3.0in]{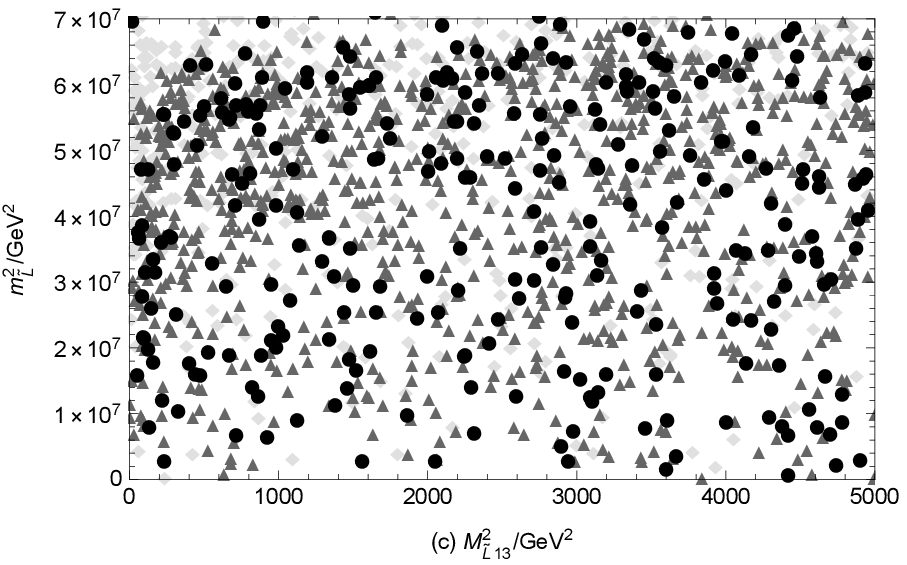}
\vspace{0.2cm}
\setlength{\unitlength}{5mm}
\centering
\includegraphics[width=3.0in]{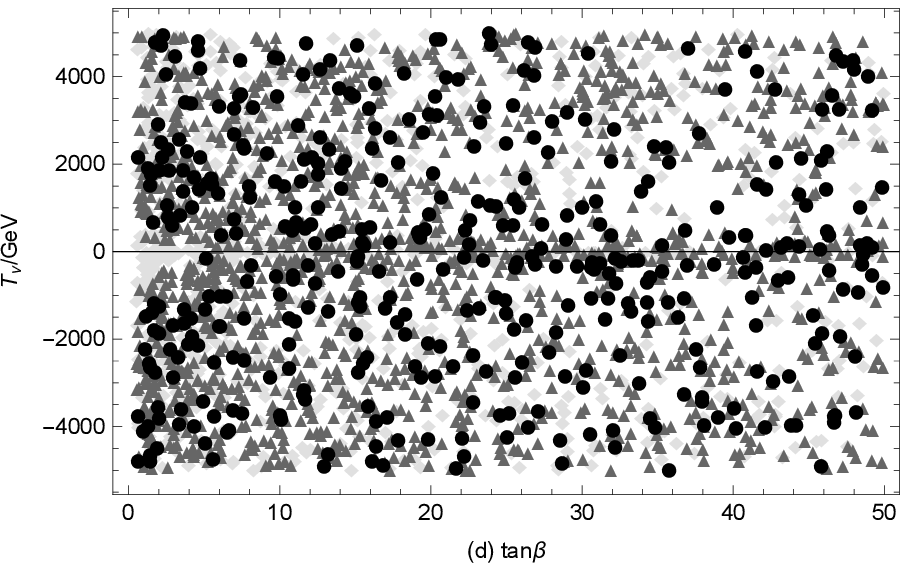}
\caption{{\label{6}}  For the scatter diagrams of the parameters below the experimental limit $Br(\tau\rightarrow{e\pi})$, different points represent the different ranges of $Br(\tau\rightarrow{e\pi})$. \textcolor{light-gray} {$\blacklozenge$} represent the range less than $0.1\times10^{-8}$. \textcolor{dark-gray}{$\blacktriangle$} represent the range of $0.1\times10^{-8}$ to $4.0\times10^{-8}$, and $\bullet$ represent the range of $4.0\times10^{-8}$ to $8.0\times10^{-8}$.}
\end{figure}

We analyze $\tau\rightarrow{e\pi}$, $\tau\rightarrow{e\eta}$ and $\tau\rightarrow{e\eta^\prime}$
 to study the possibility of lepton flavor violation.
 Fig.\ref {6} is based on the Table VI. The branching ratio of $\tau\rightarrow{e\pi}$ process is denoted by: \textcolor{light-gray} {$\blacklozenge$} (0$<Br(\tau\rightarrow{e\pi})<0.1\times10^{-8}$), \textcolor{dark-gray}{$\blacktriangle$} ($0.1\times10^{-8}$ $\leq Br(\tau\rightarrow{e\pi})<4.0\times10^{-8}$) and $\bullet$ ($Br(\tau\rightarrow{e\pi})$ from $4.0\times10^{-8}$ to $8.0\times10^{-8}$).
  The Fig.\ref {6}(a) shows the effect from $\tan\beta$ and $m_{\tilde{L}}^2$.
   Most points are concentrated in the upper left corner and distributed near the ordinate
   and $m_{\tilde{L}}^2=6\times10^7{\rm GeV}^2$.  \textcolor{light-gray} {$\blacklozenge$} are on the innermost side of the whole region.
   \textcolor{dark-gray}{$\blacktriangle$} are in the middle and $\bullet$ are on the outermost side. The trajectories of the three types of points are increasing functions.
  In Fig.\ref {6}(b), we analyze the effect of $T_{\nu}$ and $m_{\tilde{L}}^2$ on $Br(\tau\rightarrow{e\pi})$. Most of the three types points are at $5\times10^7~{\rm GeV^2} <m_{\tilde{L}}^2<7\times10^7~\rm GeV^2$ and -1000 GeV$<T_{\nu}<1000$ GeV. There are obvious stratification: As the value of $m_{\tilde{L}}^2$ becomes smaller and smaller, it is distributed in \textcolor{light-gray} {$\blacklozenge$}, \textcolor{dark-gray}{$\blacktriangle$} and $\bullet$ in turn.
  In the Fig.\ref {6}(c) with two axes $m_{\tilde{L}}^2$ versus $M_{\tilde{L}13}^2$, most of
 the points concentrate in $4\times10^7~{\rm GeV^2}<m_{\tilde{L}}^2<7\times10^7~{\rm GeV^2}$
 and $0<M_{\tilde{L}13}^2<500~{\rm GeV^2}$.
 It is worth noting that \textcolor{light-gray} {$\blacklozenge$} are obviously distributed in the upper left corner.
  This indicates that when the value of $M_{\tilde{L}13}^2$ is smaller and the value of $m_{\tilde{L}}^2$ is larger, it is easier to meet the experimental limit.
 The effect from $T_{\nu}$ and $\tan\beta$ is shown in Fig.\ref {6}(d).
 All points are almost symmetrically distributed with $T_{\nu}=0$ as the axis. Taking the upper half as an example, the $\bullet$, \textcolor{dark-gray}{$\blacktriangle$} and \textcolor{light-gray} {$\blacklozenge$} like curves that decrease and gradually approach $T_{\nu}=0$, and the values decrease in turn.

\begin{table*}
\caption{Scanning parameters for Fig.{\ref {6}}}\label{VI}
\begin{tabular*}{\textwidth}{@{\extracolsep{\fill}}lllll@{}}
\hline
Parameters&Min&Max\\
\hline
$m^2_{\tilde{\nu}}/\rm GeV^2$&$5\times10^5$&$5\times10^6$\\
$m_{\tilde{L}}^2/\rm GeV^2$&$2\times10^5$&$7\times10^7$\\
\hline
\end{tabular*}
\end{table*}

\begin{figure}[h]
\setlength{\unitlength}{5mm}
\centering
\includegraphics[width=3.0in]{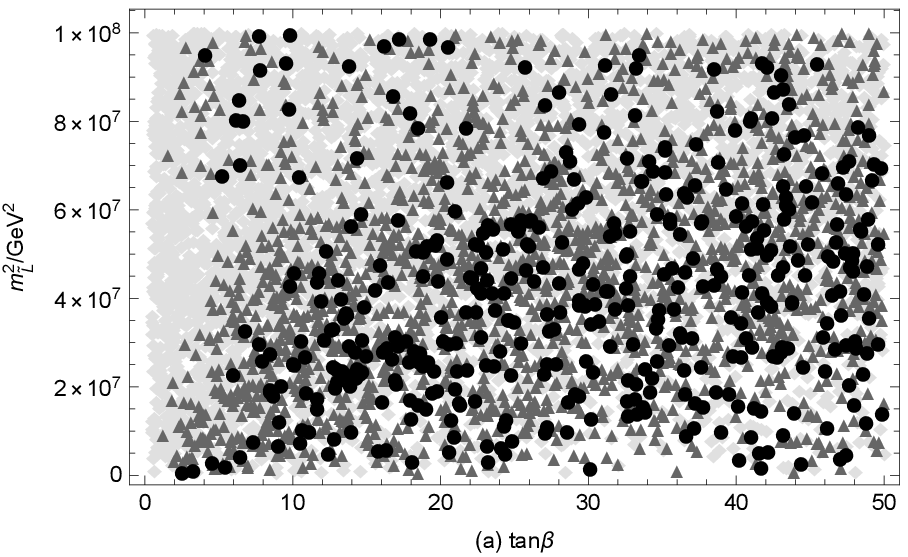}
\vspace{0.2cm}
\setlength{\unitlength}{5mm}
\centering
\includegraphics[width=3.0in]{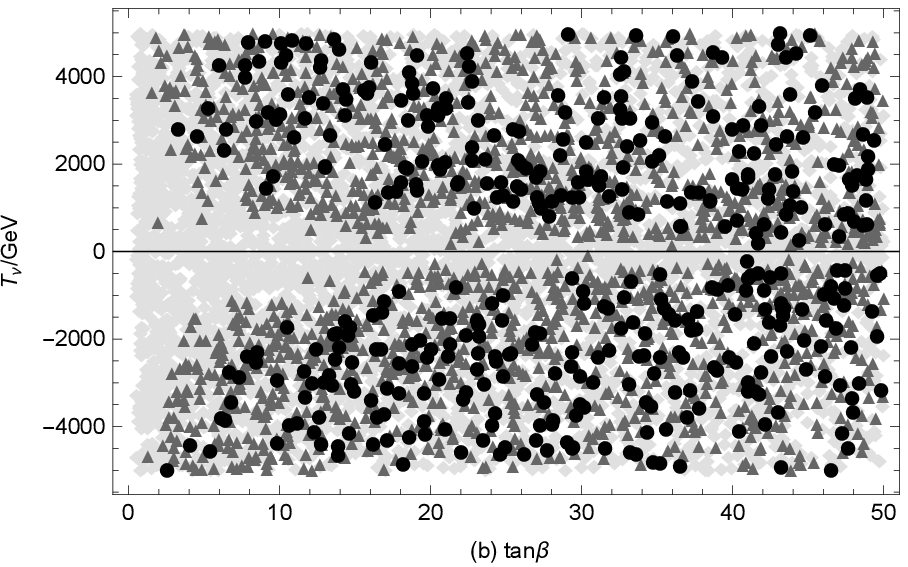}
\vspace{0.2cm}
\setlength{\unitlength}{5mm}
\centering
\includegraphics[width=3.0in]{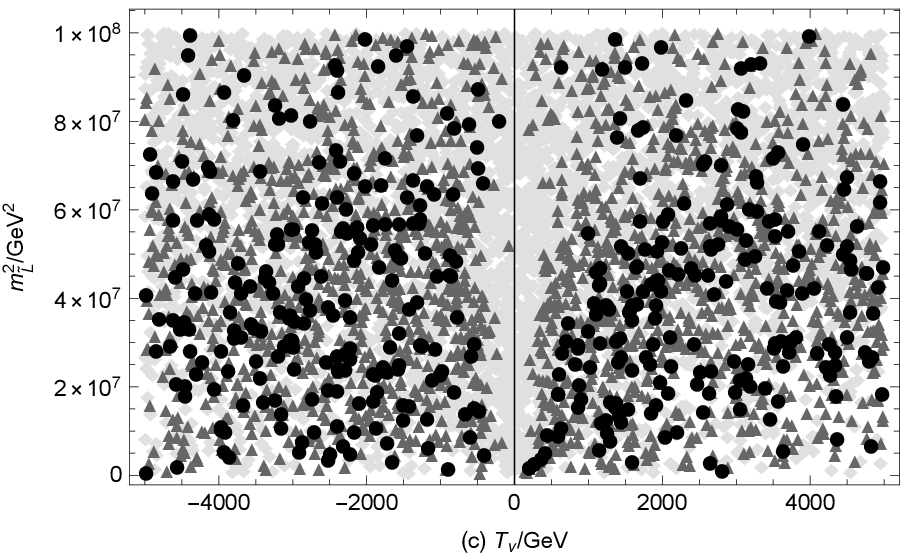}
\vspace{0.2cm}
\setlength{\unitlength}{5mm}
\centering
\includegraphics[width=3.0in]{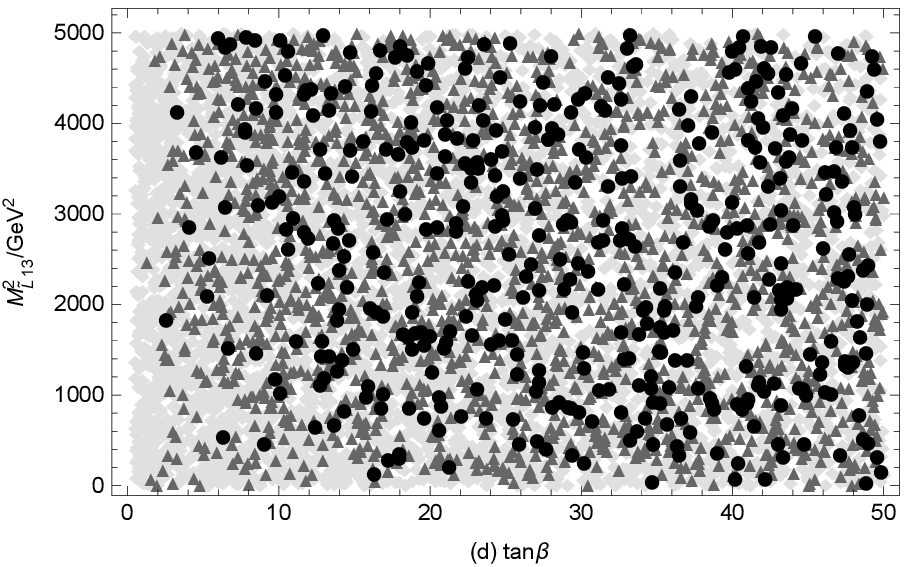}
\caption{ {\label{7}}Scatter points on the parameters under the limit of $Br(\tau\rightarrow{e\eta})$ less than $9.2\times10^{-8}$.  \textcolor{light-gray} {$\blacklozenge$}, \textcolor{dark-gray}{$\blacktriangle$} and $\bullet$ represent 0 $<Br(\tau\rightarrow{e\eta})<0.1\times10^{-8}$, $0.1\times10^{-8}\leq Br(\tau\rightarrow{e\eta})<6.0\times10^{-8}$ and $6.0\times10^{-8}\leq Br(\tau\rightarrow{e\eta})<9.2\times10^{-8}$ respectively.}
\end{figure}

Next, we scatter points on $\tau\rightarrow{e\eta}$ in Fig.\ref {7} with the parameters in the Table \ref {VII}.
  \textcolor{light-gray} {$\blacklozenge$}, \textcolor{dark-gray}{$\blacktriangle$} and $\bullet$ indicate the range 0 $<Br(\tau\rightarrow{e\eta})<0.1\times10^{-8}$, $0.1\times10^{-8} \leq Br(\tau\rightarrow{e\eta})<6.0\times10^{-8}$ and $6.0\times10^{-8}\leq Br(\tau\rightarrow{e\eta})<9.2\times10^{-8}$  respectively.
   In Fig.\ref {7}(a), we show the points in the plane of $\tan\beta$ and $m_{\tilde{L}}^2$.
  \textcolor{light-gray} {$\blacklozenge$} basically occupy the whole parameter space, which are more obvious in the upper left corner.
$\bullet$ are mainly distributed in the lower right corner, and \textcolor{dark-gray}{$\blacktriangle$} are mostly concentrated in the middle part.
 We show $\tan\beta$ versus $T_{\nu}$ in Fig.\ref {7}(b).
 The whole diagram is almost symmetrical about $T_{\nu}=0$ and above the $T_\nu=0$ the overall trend is downward. \textcolor{light-gray} {$\blacklozenge$} are mainly in $0.5<\tan\beta<10$
 and on both sides of the $T_{\nu}=0$ axis. Then the outer side
 are \textcolor{dark-gray}{$\blacktriangle$}, and the outermost side are $\bullet$.
  Next, in Fig.\ref {7}(c), we show the relationship between $T_{\nu}$ and $m_{\tilde{L}}^2$. Most of the points are basically symmetrical about axis $T_{\nu}=0$. \textcolor{light-gray} {$\blacklozenge$} mainly concentrate near $T_{\nu}=0$.
 \textcolor{dark-gray}{$\blacktriangle$} and $\bullet$ are mainly concentrated in $0<m_{\tilde{L}}^2<6\times10^7~\rm GeV^2$ and gradually away from $T_{\nu}=0$. Finally, we analyze the impact of $\tan\beta$ and $M_{\tilde{L}13}^2$ on $Br(\tau\rightarrow\mu\eta)$ in Fig.\ref {7}(d). \textcolor{light-gray} {$\blacklozenge$} mainly concentrate in $\tan\beta$ range from 0.5 to 3 and \textcolor{dark-gray}{$\blacktriangle$} concentrate in area $3<\tan\beta<10$.

\begin{table*}
\caption{Scanning parameters for Fig.12 and Fig.13}\label{VII}
\begin{tabular*}{\textwidth}{@{\extracolsep{\fill}}lllll@{}}
\hline
Parameters&Min&Max\\
\hline
$m^2_{\tilde{\nu}}/\rm GeV^2$&$5\times10^5$&5$\times10^7$\\
$m_{\tilde{L}}^2/\rm GeV^2$&$5\times10^5$&$10^8$\\
\hline
\end{tabular*}
\end{table*}

\begin{figure}[h]
\setlength{\unitlength}{5mm}
\centering
\includegraphics[width=3.0in]{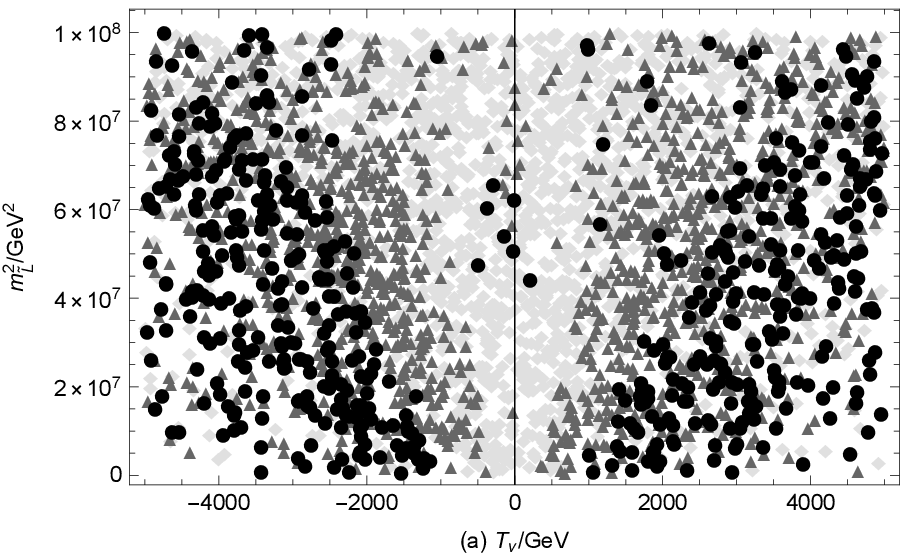}
\vspace{0.2cm}
\setlength{\unitlength}{5mm}
\centering
\includegraphics[width=3.0in]{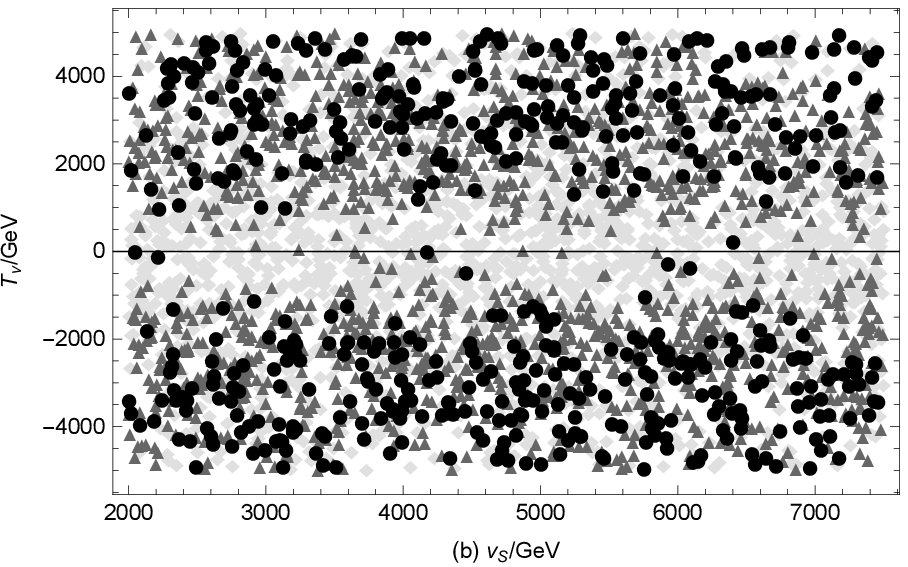}
\caption{ {\label{8}}Considering the limit of $Br(\tau\rightarrow{e\eta^\prime})$, the diagrams are plotted, where \textcolor{light-gray} {$\blacklozenge$} represent 0$<Br(\tau\rightarrow{e\eta^\prime})<1.0\times10^{-8}$, \textcolor{dark-gray}{$\blacktriangle$} represent $1.0\times10^{-8}\leq Br(\tau\rightarrow{e\eta^\prime})<8.0\times10^{-8}$ and $8.0\times10^{-8}\leq Br(\tau\rightarrow{e\eta^\prime})<1.6\times10^{-7}$ are represented by $\bullet$.}
\end{figure}

The drawing of Fig.\ref {8} is similar to the previous one. We draw the process of $\tau\rightarrow{e\eta^\prime}$ according to table \ref {VII}. \textcolor{light-gray} {$\blacklozenge$} indicate that the value of $Br(\tau\rightarrow{e\eta^\prime})$ is from 0 to $1.0\times10^{-8}$.
 \textcolor{dark-gray}{$\blacktriangle$} indicate a range of
 $1.0\times10^{-8}\leq Br(\tau\rightarrow\mu\eta^\prime)<8.0\times10^{-8}$ and $\bullet$
 are $8.0\times10^{-8}\leq Br(\tau\rightarrow\mu\eta^\prime)<1.6\times10^{-7}$.
 We describe the effect of $m_{\tilde{L}}^2$ and $T_\nu$
on $Br(\tau\rightarrow{e\eta^\prime})$ in Fig.\ref {8}(a). We can easily see that \textcolor{light-gray} {$\blacklozenge$}, \textcolor{dark-gray}{$\blacktriangle$} and $\bullet$ all have the same downward trend, which are symmetrical about $T_\nu=0$. \textcolor{light-gray} {$\blacklozenge$} are distributed on both sides of $T_\nu=0$, followed by \textcolor{dark-gray}{$\blacktriangle$}, and the outermost layer are $\bullet$. Interestingly, the smaller the value of $m_{\tilde{L}}^2$, the closer the trend of \textcolor{dark-gray}{$\blacktriangle$} and $\bullet$ are to $T_\nu=0$.
We show $T_{\nu}$ versus $v_S$ in the Fig.\ref {8}(b). The points have obvious stratification and are generally symmetrical about $T_{\nu}=0$.
Taking the part of  $T_{\nu}>0$ as an example, numerically, $\bullet$ are the largest, followed by \textcolor{dark-gray}{$\blacktriangle$}  and the smallest are \textcolor{light-gray} {$\blacklozenge$}. In other words, on both sides of $T_{\nu}=0$, there are \textcolor{light-gray} {$\blacklozenge$}, \textcolor{dark-gray}{$\blacktriangle$} and $\bullet$ from inside to outside.

\section{discussion and conclusion}

The $U(1)_X$SSM is the U(1) extension of MSSM, and its local
gauge group is $SU(3)_C\times SU(2)_L \times U(1)_Y\times U(1)_X$.
Comparing with MSSM, $U(1)_X$SSM has new superfields including righ-handed neutrinos and three Higgs superfields $\hat{\eta},~\hat{\bar{\eta}},~\hat{S}$, which leads to several advantages as introduced in the introduction.
 This model possesses more LFV sources, which can give considerable
contributions to the LFV processes $\tau\rightarrow Pl$. Considering
the constraints from neutrino experiments, $\tau\rightarrow\mu\gamma$, $\tau\rightarrow e\gamma$, $\mu\rightarrow e \gamma$ and $\mu\rightarrow eee$, we study
the one loop diagrams which include self-energy diagrams, triangle diagrams and box diagrams for $\tau\rightarrow Pl$.
In the numerical calculation, we scan large parameter spaces and make rich numerical results.
As a result, many diagrams of the numerical results are plotted.

In this work, we use the effective
Lagrangian method to calculate and analyze the LFV decays $l_j\rightarrow l_i\gamma$, $\mu\rightarrow eee$ and $\tau\rightarrow{Pl}$
in the framework of the $U(1)_{X}$SSM model. We scan a large number
of parameters to find the possibility of LFV. After comparison and analysis,
$\tan\beta$, $T_{\nu}$, $M^2_{\tilde{\nu}13}$, $M^2_{\tilde{L}23}$ and $M^2_{\tilde{L}13}$
are sensitive parameters that have obvious impacts on the results.
$v_S$ and $M_2$ are insensitive parameters. In the whole, the non-diagonal elements which correspond to the generations of the initial lepton and final lepton
are main sensitive parameters and LFV sources.
 Most parameters can break the upper limit of the experiment and provide new ideas for finding NP.

{\bf Acknowledgments}

This work is supported by National Natural Science Foundation of China (NNSFC)
	(No. 11535002, No. 11705045), Natural Science Foundation of Hebei Province
	(A2020201002). Post-graduate's Innovation Fund Project of Hebei University
    (HBU2022ss028).

\appendix
\section{mass matrix}\label{A1}

The mass squared matrix for slepton with the basis $(\tilde{e}_L, \tilde{e}_R)$ is diagonalized by $Z^E$ through the
formula $Z^E m^2_{\tilde{e}} Z^{E,\dagger} = m^{diag}_{2,\tilde{e}}$,
\begin{equation}
m^2_{\tilde{e}} = \left(
\begin{array}{cc}
m_{\tilde{e}_L\tilde{e}_L^*} &\frac{1}{2} \Big(\sqrt{2} v_d T_{e}^{\dagger}  - v_u \Big({\lambda}_{H} v_S  + \sqrt{2} \mu \Big)Y_{e}^{\dagger} \Big)\\
\frac{1}{2} \Big(\sqrt{2} v_d T_e  - v_u Y_e \Big(\sqrt{2} \mu^*  + v_S {\lambda}_{H}^* \Big)\Big) &m_{\tilde{e}_R\tilde{e}_R^*}\end{array}
\right),
 \end{equation}
\begin{eqnarray}
&&m_{\tilde{e}_L\tilde{e}_L^*} = M_{\tilde{L}}^2+\frac{1}{8} \Big((g_{1}^{2} + g_{Y X}^{2}
+ g_{Y X} g_{X} -g_2^2)(v_{d}^{2}- v_{u}^{2})+ 2 g_{Y X} g_{X}( v_{\eta}^{2}- v_{\bar{\eta}}^{2}
)
\Big)+\frac{1}{2} v_{d}^{2} {Y_{e}^{\dagger}  Y_e} ,\nonumber\\&&
m_{\tilde{e}_R\tilde{e}_R^*} = M_{\tilde{E}}^2-\frac{1}{8}  \Big([2(g_{1}^{2} + g_{Y X}^{2})+3g_{Y X} g_{X}+g_{X}^{2}]
( v_{d}^{2}- v_{u}^{2})\nonumber\\&&\hspace{1.6cm}+(4g_{Y X} g_{X}+2g_{X}^{2})(v_{\eta}^{2}- v_{\bar{\eta}}^{2})
\Big)+\frac{1}{2} v_{d}^{2} {Y_e  Y_{e}^{\dagger}}.
\end{eqnarray}

The mass matrix for neutralino in the basis $(\lambda_{\tilde{B}}, \tilde{W}^0, \tilde{H}_d^0, \tilde{H}_u^0,
\lambda_{\tilde{X}}, \tilde{\eta}, \tilde{\bar{\eta}}, \tilde{s}) $ is

\begin{equation}
m_{\tilde{\chi}^0} = \left(
\begin{array}{cccccccc}
M_1 &0 &-\frac{g_1}{2}v_d &\frac{g_1}{2}v_u &{M}_{B B'} &0  &0  &0\\
0 &M_2 &\frac{1}{2} g_2 v_d  &-\frac{1}{2} g_2 v_u  &0 &0 &0 &0\\
-\frac{g_1}{2}v_d &\frac{1}{2} g_2 v_d  &0
&m_{\tilde{H}_d^0\tilde{H}_u^0} &m_{\tilde{H}_d^0\lambda_{\tilde{X}}} &0 &0 & - \frac{{\lambda}_{H} v_u}{\sqrt{2}}\\
\frac{g_1}{2}v_u &-\frac{1}{2} g_2 v_u  &m_{\tilde{H}_d^0\tilde{H}_u^0} &0 &m_{\tilde{H}_u^0\lambda_{\tilde{X}}} &0 &0 &- \frac{{\lambda}_{H} v_d}{\sqrt{2}}\\
{M}_{B B'} &0 &m_{\tilde{H}_d^0\lambda_{\tilde{X}}} &m_{\tilde{H}_u^0\lambda_{\tilde{X}}} &{M}_{BL} &- g_{X} v_{\eta}  &g_{X} v_{\bar{\eta}}  &0\\
0  &0 &0 &0 &- g_{X} v_{\eta}  &0 &\frac{1}{\sqrt{2}} {\lambda}_{C} v_S  &\frac{1}{\sqrt{2}} {\lambda}_{C} v_{\bar{\eta}} \\
0  &0 &0 &0 &g_{X} v_{\bar{\eta}}  &\frac{1}{\sqrt{2}} {\lambda}_{C} v_S  &0 &\frac{1}{\sqrt{2}} {\lambda}_{C} v_{\eta} \\
0 &0 & - \frac{{\lambda}_{H} v_u}{\sqrt{2}} &- \frac{{\lambda}_{H} v_d}{\sqrt{2}} &0 &\frac{1}{\sqrt{2}} {\lambda}_{C} v_{\bar{\eta}}
 &\frac{1}{\sqrt{2}} {\lambda}_{C} v_{\eta}  &m_{\tilde{s}\tilde{s}}\end{array}
\right),\label{neutralino}
 \end{equation}

\begin{eqnarray}
&& m_{\tilde{H}_d^0\tilde{H}_u^0} = - \frac{1}{\sqrt{2}} {\lambda}_{H} v_S  - \mu ,~~~~~~~
m_{\tilde{H}_d^0\lambda_{\tilde{X}}} = -\frac{1}{2} \Big(g_{Y X} + g_{X}\Big)v_d, \nonumber\\&&
m_{\tilde{H}_u^0\lambda_{\tilde{X}}} = \frac{1}{2} \Big(g_{Y X} + g_{X}\Big)v_u
 ,~~~~~~~~~~~~
m_{\tilde{s}\tilde{s}} = 2 M_S  + \sqrt{2} \kappa v_S.\label{neutralino1}
\end{eqnarray}
This matrix is diagonalized by $Z^N$,
\begin{equation}
Z^{N*} m_{\tilde{\chi}^0} Z^{N{\dagger}} = m^{diag}_{\tilde{\chi}^0}.
\end{equation}
\section{Neutrino Mixing}\label{A2}
 We obtain the eigenvalues of the $3\times3$ matrix $\mathcal{H}$
\begin{eqnarray}
&&m^2_1=\frac{a}{3}-\frac{1}{3}p(\cos\phi+\sqrt{3}\sin\phi),\nonumber\\
&&m^2_2=\frac{a}{3}-\frac{1}{3}p(\cos\phi-\sqrt{3}\sin\phi),\nonumber\\
&&m^2_3=\frac{a}{3}+\frac{2}{3}p\cos\phi.
\end{eqnarray}
In order to express the formulae concisely, we define the
notations
\begin{eqnarray}
&&p=\sqrt{a^2-3b},\nonumber\\
&&\phi=\frac{1}{3}\arccos(\frac{1}{p^3}(a^3-\frac{9}{2}ab+\frac{27}{2}c)),
\end{eqnarray}
with
\begin{eqnarray}
&&a=Tr(\mathcal{H}),\nonumber\\
&&b=\mathcal{H}_{11}\mathcal{H}_{22}+\mathcal{H}_{11}\mathcal{H}_{33}+\mathcal{H}_{22}\mathcal{H}_{33}
-\mathcal{H}_{12}^2-\mathcal{H}_{13}^2-\mathcal{H}_{23}^2,\nonumber\\
&&c=Det(\mathcal{H}).
\end{eqnarray}

To explain the experimental data from neutrino oscillations, we arrange that $m_1^2<m_2^2<m_3^2$. In the case of 3-neutrino mixing,  we consider the spectrum with normal ordering(NO):
\begin{eqnarray}
&&m_{\nu1}<m_{\nu2}<m_{\nu3},\nonumber\\
&&m_{\nu1}^2=m_1^2,~~m_{\nu2}^2=m_2^2,~~m_{\nu3}^2=m_3^2,\nonumber\\
&&\Delta{m_{\bigodot}^2}=m_{\nu2}^2-m_{\nu1}^2=\frac{2}{\sqrt{3}}p\sin\phi>0,\nonumber\\
&&\Delta{m_{A}^2}=m_{\nu3}^2-m_{\nu1}^2=p(\cos\phi+\frac{1}{\sqrt{3}}\sin\phi)>0.
\end{eqnarray}

The normalized eigenvectors for the mass squared matrix $\mathcal{H}$ are given as
\begin{eqnarray}
\left(
\begin{array}{c}
(U_{\nu})_{11}\\
(U_{\nu})_{21}\\
(U_{\nu})_{31}\\
\end{array}
\right)
=\frac{1}{\sqrt{|X_1|^2+|Y_1|^2+|Z_1|^2}}
\left(
\begin{array}{c}
X_1\\
Y_1\\
Z_1\\
\end{array}
\right),\nonumber\\
\left(
\begin{array}{c}
(U_{\nu})_{12}\\
(U_{\nu})_{22}\\
(U_{\nu})_{32}\\
\end{array}
\right)
=\frac{1}{\sqrt{|X_2|^2+|Y_2|^2+|Z_2|^2}}
\left(
\begin{array}{c}
X_2\\
Y_2\\
Z_2\
\end{array}
\right),\nonumber\\
\left(
\begin{array}{c}
(U_{\nu})_{13}\\
(U_{\nu})_{23}\\
(U_{\nu})_{33}\\
\end{array}
\right)
=\frac{1}{\sqrt{|X_3|^2+|Y_3|^2+|Z_3|^2}}
\left(
\begin{array}{c}
X_3\\
Y_3\\
Z_3\\
\end{array}
\right),
\end{eqnarray}
with
\begin{eqnarray}
&&X_1=(\mathcal{H}_{22}-m_{\nu1}^2)(\mathcal{H}_{33}-m_{\nu1}^2)-\mathcal{H}_{23}^2,\nonumber\\
&&Y_1=\mathcal{H}_{13}\mathcal{H}_{23}-\mathcal{H}_{12}(\mathcal{H}_{33}-m_{\nu1}^2),\nonumber\\
&&Z_1=\mathcal{H}_{12}\mathcal{H}_{23}-\mathcal{H}_{13}(\mathcal{H}_{22}-m_{\nu1}^2),\nonumber\\
&&X_2=\mathcal{H}_{13}\mathcal{H}_{23}-\mathcal{H}_{12}(\mathcal{H}_{33}-m_{\nu2}^2),\nonumber\\
&&Y_2=(\mathcal{H}_{11}-m_{\nu2}^2)(\mathcal{H}_{33}-m_{\nu2}^2)-\mathcal{H}_{13}^2,\nonumber\\
&&Z_2=\mathcal{H}_{12}\mathcal{H}_{13}-\mathcal{H}_{23}(\mathcal{H}_{11}-m_{\nu2}^2),\nonumber\\
&&X_3=\mathcal{H}_{12}\mathcal{H}_{23}-\mathcal{H}_{13}(\mathcal{H}_{22}-m_{\nu3}^2),\nonumber\\
&&Y_3=\mathcal{H}_{12}\mathcal{H}_{13}-\mathcal{H}_{23}(\mathcal{H}_{11}-m_{\nu3}^2),\nonumber\\
&&Z_3=(\mathcal{H}_{11}-m_{\nu3}^2)(\mathcal{H}_{22}-m_{\nu3}^2)-\mathcal{H}_{12}^2.
\end{eqnarray}

 In a general way, we define the mixing angles among three tiny neutrinos as
\begin{eqnarray}
&&\sin\theta_{13}=|(U_\nu)_{13}|,~~~~~~~~~~~~\cos\theta_{13}=\sqrt{1-|(U_\nu)_{13}|^2},\nonumber\\
&&\sin\theta_{23}=\frac{|(U_\nu)_{23}|}{\sqrt{1-|(U_\nu)_{13}|^2}},~~\cos\theta_{23}=\frac{|(U_\nu)_{33}|}{\sqrt{1-|(U_\nu)_{13}|^2}},\nonumber\\
&&\sin\theta_{12}=\frac{|(U_\nu)_{12}|}{\sqrt{1-|(U_\nu)_{13}|^2}},~~\cos\theta_{12}=\frac{|(U_\nu)_{11}|}{\sqrt{1-|(U_\nu)_{13}|^2}}.
\end{eqnarray}

\end{document}